\documentclass{article}%%%%where rstrans is the template name

\usepackage[margin=1in]{geometry}

\usepackage{breakcites}

\usepackage{amsmath}
\usepackage{times}
\usepackage[english]{babel}
\usepackage[latin1]{inputenc}
\usepackage{graphicx}
\usepackage{xspace}
\usepackage{lscape}
\usepackage{indentfirst}
\usepackage{graphicx}
\usepackage{amsmath}

\usepackage{enumitem}

\usepackage{xcolor} % (COLORI) usare per evidenziare correzioni:
\usepackage{amsthm}
\usepackage{url}

\newtheoremstyle{ourstyle}% name
  {4mm}%      Space above, empty = `usual value'
  {3pt}%      Space below
  {\upshape}% Body font
  {}%         Indent amount (empty = no indent, \parindent = para indent)
  {\bfseries}% Thm head font
  {.}%        Punctuation after thm head
  {.5em}%     Space after thm head: " " = normal interword space;
  {}% Thm head spec

\theoremstyle{ourstyle}

\newtheorem{thm}{Theorem}
\newtheorem{lmm}{Lemma}
\newtheorem{coroll}{Corollary}
\newtheorem{assumption}{Assumption}
\newtheorem{assumptionrelaxed}{Relaxed Assumption}
\newtheorem{previewassumption}{Assumption}
\newtheorem{previewassumptionrelaxed}{Relaxed Assumption}
\newtheorem{postulate}{Postulate}
\newtheorem*{definition}{Definition}
\newtheorem*{comment}{Remark}
\newcounter{footnumber}
\newcommand{\dimostrazione}[3]{{\setcounter{footnumber}{\value{footnote}}\addtocounter{footnumber}{
1}\setcounter{footnote}{\value{footnumber}}The proof is  in Footnote \thefootnumber.\footnotetext{\textbf{Proof of #1 \ref{#2}.\ }#3\qed\\ }\setcounter{footnumber}{\thefootnote}}}

\makeatletter

\renewcommand\normalsize{%
   \@setfontsize\normalsize\@ixpt\@xiipt
   \abovedisplayskip 6\p@ \@plus2\p@ \@minus2\p@
   \abovedisplayshortskip \z@ \@plus3\p@
   \belowdisplayshortskip 6\p@ \@plus3\p@ \@minus3\p@
   \belowdisplayskip \abovedisplayskip
   \let\@listi\@listI}

\renewcommand\footnotesize{%
   \@setfontsize\footnotesize\@ixpt{9}%
   \abovedisplayskip 3\p@ \@plus\p@ \@minus2\p@
   \abovedisplayshortskip \z@ \@plus\p@
   \belowdisplayshortskip 3\p@ \@plus\p@ \@minus2\p@
   \def\@listi{\leftmargin\leftmargini
               \topsep 3\p@ \@plus\p@ \@minus\p@
               \parsep 2\p@ \@plus\p@ \@minus\p@
               \itemsep \parsep}%
   \belowdisplayskip \abovedisplayskip}

\makeatother

\newcommand{\virgolettedue}[1]{{``#1''}}
\newcommand{\W}[3]{{{#1}_#2\xrightarrow{\rm w} {#1}_#3}}

\newcommand{\Wcomp}[2]{{{#1}_1#2_{\rm se1}\xrightarrow{\rm w} #1_2#2_{\rm se2}}}
\newcommand{\Wcomprev}[2]{{#1_1#2_{\rm se1}\xrightarrow{\rm wrev}#1_2#2_{\rm se2rev}}}
\newcommand{\RA}[1]{{\pmb{\textsl{\textsf{R}}}^A_{#1}}}
\newcommand{\RB}[1]{{\pmb{\textsl{\textsf{R}}}^B_{#1}}}
\newcommand{\RC}[1]{{\pmb{\textsl{\textsf{R}}}^C_{#1}}}
\newcommand{\RD}[1]{{\pmb{\textsl{\textsf{R}}}^D_{#1}}}

\newcommand{\spazio}{{\vskip 3pt\noindent}}

\newcommand{\Hil}{{\mathcal H}}
\newcommand{\Boltz}{ k}
\newcommand{\Tr}{{\rm Tr}}
\newcommand{\ih}{\rotatebox[origin=c]{180}{$h$}}

\title{New definitions of thermodynamic temperature and entropy not based on the concepts of heat and thermal reservoir}

\author{Gian Paolo Beretta$^1$ and Enzo Zanchini$^2$\\  $^1$ Universit\'a di Brescia, Italy\\ $^2$ Universit\`{a} di Bologna, Italy}

\date{published in Atti della Accademia Peloritana dei Pericolanti, Vol. 97(S1), A1, 1-28 (2019)\\
	\url{http://dx.doi.org/10.1478/AAPP.97S1A1}}

\begin{document}
	\maketitle

\begin{abstract}
From a new rigorous formulation of the general axiomatic foundations of thermodynamics we derive an operational definition of entropy that responds to the emergent need in many technological frameworks to understand and deploy thermodynamic entropy well beyond the traditional realm of equilibrium states of macroscopic systems. The new treatment starts from a previously developed set of carefully worded operational definitions for all the necessary basic concepts, and is not based on the traditional ones of \virgolettedue{heat} and of \virgolettedue{thermal reservoir}. It is achieved in three steps. First, a new definition of thermodynamic temperature is stated, for any stable equilibrium state. Then, by employing this definition, a measurement procedure is developed which defines uniquely the property entropy in a broad domain of states, which could include \emph{in principle}, even some non-equilibrium states of few-particle systems, provided they are separable and uncorrelated. Finally, the domain of validity of the definition is extended, possibly to every state of every system, by a different procedure, based on the preceding one, which associates a range of entropy values to any state not included in the previous domain. The principle of entropy non-decrease and the additivity of entropy are proved in both the domains considered.
\end{abstract}

%\keyword{definition of entropy; second law of thermodynamics;  non-equilibrium entropy; few-particle thermodynamics}

\section{Introduction}

Thermodynamic entropy plays a crucial role in the development of the physical foundations of a variety of emerging technologies --- nanomaterials, small-scale hydrodynamics, chemical kinetics for energy and environmental engineering and biotechnologies, electrochemistry, quantum entanglement in quantum information, non-equilibrium bulk and interface phenomena,  etc.\ ---  which require a clear understanding of the meaning and role of thermodynamic entropy beyond the traditional equilibrium and macroscopic realms, well into the non-equilibrium and few-particle domains currently being explored very actively in  many fields of science and technology (see, e.g., Refs.\ \cite{Horodecki13,Skrzypczyk14,Brandao13,Verley12} for recent attempts to extend thermodynamics to nonequilibrium states and individual quantum systems).

In traditional treatments of thermodynamics (see, e.g. Refs.\ \cite{Fermi37, Pippard57, Zemansky68}), the definitions of thermodynamic temperature and of entropy are based on the concepts of \emph{heat} and of \emph{thermal reservoir}. Usually, heat is not defined rigorously. For instance, in his lectures on physics, Feynman \cite{Feynman63} describes heat as one of several different forms of energy related to the jiggling motion of particles; in this picture, heat appears as a transfer of kinetic energy and the difference between heat and work is not clarified. Landau \cite{Landau80} defines heat as the part of an energy change of a body that is not due to work done on it. However, there are interactions between systems which are neither heat nor work, such as, for instance, exchanges of radiation between systems in nonequilibrium states. Guggenheim \cite{Guggenheim67} defines heat as an exchange of energy that differs from work and is determined by a temperature difference. Keenan \cite{Keenan41} defines heat as the energy transferred from one system to a second system at lower temperature, by virtue of the temperature difference, when the two are brought into communication. These definitions do not describe clearly the phenomena which occur at the interface between the interacting systems; moreover, they require a previous definition of \emph{empirical temperature}, a concept which, in turn, is usually not defined rigorously. Another drawback of the employment of heat in the definition of entropy is the following: since heat, when properly defined, requires the existence of subsystems in stable equilibrium at the boundary between the interacting systems, a definition of entropy based on heat can hold, at most, in the domain of \emph{local equilibrium} states.

An alternative method for the axiomatization of thermodynamics was developed at MIT by Hatsopoulos and Keenan \cite{HK} and by Gyftopoulos and Beretta \cite{GB}. The main progress obtained in these references, with respect to the traditional treatments, is a more general definition of entropy --- not based on the heuristic notions of empirical temperature and heat, and not restricted  \emph{a priori}   to stable equilibrium states --- that emerges from a complete set of operational definitions of the basic concepts, such as those of system, property, state, and stable equilibrium state, and a new statement of the second law expressed as a postulate of existence, for a system with fixed composition and constraints, of  a unique stable equilibrium state for each  value of the energy.

Improvements of this method, yielding more rigorous definitions of isolated system, environment of a system and external force field, as well as a more direct definition of entropy, have been proposed over the years by the present authors \cite{Zanchini86,Zanchini88,Zanchini92,ZBsymposium,ZBIJoT, BZInTech, ZBEntropy14}.  Such constructions are important because they provide rigorous operational definitions of entropy potentially valid also in the non-equilibrium domain.   However, they still require the use of a thermal reservoir as an auxiliary system (that plays the role of an \textit{entropy meter}) in the operational procedure that defines how to measure the entropy difference between any two states of a system. As already pointed out in Ref.\ \cite[p.87]{GB}, such use of thermal reservoirs has both logical and operational drawbacks.

A thermal reservoir, when properly defined \cite{GB, ZBIJoT, BZInTech}, is a closed system $R$, contained in a fixed region of space, such that whenever $R$  is in stable equilibrium it is also in mutual stable equilibrium with a duplicate of itself, kept in any of its stable equilibrium states. Once thermodynamic temperature has been defined,  it turns out that a thermal reservoir has the same temperature in all its stable equilibrium states, independently of the value of the energy. This condition is fulfilled by the simple-system model\footnote{As defined and discussed in Ref.\ \cite[pp.263-265]{GB}, the simple-system model is appropriate for macroscopic systems with many particles, but fails for few-particle systems for which, e.g., rarefaction effects near walls cannot be neglected.} of a pure substance kept  in the range of triple-point stable equilibrium states, because within such range of states energy can be added or removed at constant volume without changing the temperature. Hence, pure substances in their triple-point ranges are good practical examples of thermal reservoirs that can be easily set up in any laboratory.

However,  the triple-point model is only an approximate description of reality, valid with exceedingly good approximation for systems with many particles of the order of one mole, but not in general, e.g., not for systems with few particles. In a fully explicit axiomatic treatment one could declare the existence of thermal reservoirs as an assumption, but then one could prove that, strictly, thermal reservoirs cannot exist.
Thus, from the strictly logical point of view, the use of the thermal reservoir in the definition of entropy  is an internal inconsistency.

Another important drawback of the use of a thermal reservoir $R$ in the measurement procedure that defines the entropy difference of two states $A_1$ and $A_2$ of a system $A$  is that the procedure \cite{GB, ZBIJoT, BZInTech,ZBEntropy14} requires to measure the energy change of the reservoir $R$  in a reversible weight process for the composite system $AR$ in which $A$ changes from state $A_1$ to state $A_2$. If system $A$ has only few particles, than the energy change of $R$ will be extremely small and hardly detectable if, as just discussed, the thermal reservoir $R$  can only be realized by means of a macroscopic system.

 The scope of the present paper is to develop new general definitions of thermodynamic temperature and thermodynamic entropy that are neither based on the concept of heat nor  on that of thermal reservoir, so that both the logical and the practical drawbacks due to the use of these concepts are removed.

A procedure which yields the definitions of  temperature and  entropy without employing the concepts of heat and of thermal reservoir was presented by Constantin Carath\'eodory in 1909 \cite{Caratheodory09}. However, his treatment is valid only for stable equilibrium states of simple systems in the sense of Ref.\ \cite[Ch.17]{GB}. The same restriction holds for some developments of Carath\'eodory's method \cite{Turner60,Landsberg61, Sears63,Giles64}, aimed at making the treatment simpler and less abstract.

Another axiomatization of thermodynamics has developed in recent years by Lieb and Yngvason \cite{LY, LY13, LY14}. Their method is based on establishing an order relation between states, denoted by the symbol $\prec $, through the concept of adiabatic accessibility: a state $Y$ is said to be adiabatically accessible from a state $X$, i.e., $X \prec Y$, if it is possible to change the state from $X$ to $Y$ by means of an adiabatic process. By introducing a suitable set of Axioms concerning the order relation $\prec$, the authors prove the existence and the \emph{essential uniqueness} \cite{LY} of entropy. While the treatment presented in Ref.\  \cite{LY} holds only for stable equilibrium states of simple systems or collections of simple systems, through the complements presented in Refs.\ \cite{LY13, LY14} the validity is extended respectively to non-equilibrium states  \cite{LY13} and, through the use of a simple system as an \emph{entropy meter}, also to non-simple systems \cite{LY14}. Since to exhibit simple-system behavior the entropy meter must be a many-particle system, when applied  to few-particle systems the definition could present the same kind of 'practical'  problems faced by our previous definitions based on the entropy meter being a thermal reservoir.

In the present paper, a set of postulates and assumptions analogous to that stated in Ref. \cite{ZBEntropy14} is employed, but here the definitions of thermodynamic temperature and thermodynamic entropy are obtained \textit{without employing the concept of thermal reservoir}. The main result of the new formulation is that by avoiding to use as entropy meter a many-particle system, we derive a rigorous and general operational definition of thermodynamic entropy which holds, potentially, also for  some   non-equilibrium states of non-simple and non-macroscopic systems.  Then, the domain of validity of the definition is extended to include possibly every state of every system by a procedure, similar to that developed in \cite{LY13}, which associates a range of entropy values to any state not included in the previous domain.

The potential applicability to non-equilibrium states is a relevant feature in the framework of the fast growing field of non-equilibrium thermodynamics (see, e.g., Ref. \cite{Kjelstrup08}), where research advances seem to substantiate from many perspectives the validity of a general principle of maximum entropy production \cite{Gheorghiu01,Gheorghiu01add,Martyushev06,Beretta09,Beretta14}. It is also
relevant in the framework of the recently growing field of  thermodynamics in the quantum regime, where much discussion about the microscopic foundations of thermodynamics  is still taking place (see, e.g., Ref.\ \cite{Maddox85,%HG1,HG2,HG3,HG4,Beretta81,
Beretta86,
Hatsopoulos08,Bennett08,Lloyd08,Maccone11,BerettaEPL,Brandao15,CanoAndrade,Weilenmann,Horodecki13,Skrzypczyk14,Brandao13,Verley12}).

The definition of entropy presented here is complementary to that developed by Lieb and Yngvason: indeed, while Refs.\ \cite{LY, LY13, LY14} are focused on the proof of existence and essential uniqueness of an entropy function which is additive and fulfils the principle of entropy nondecrease, the present treatment identifies a general measurement procedure suitable to determine the entropy values.   The principle of entropy nondecrease and the additivity of entropy are then proved as consequences of the definition.

In order to focus immediately on the construction of the new general definition of entropy, we keep to a minimum the discussion of the preliminary concepts. Instead, we provide in footnotes full proofs of the lemmas, theorems, and corollaries.

\section{\label{BasicDefinitions}Summary of basic preliminary definitions}
 In this section, we very briefly  summarize the definitions of terms and preliminary concepts that we will use in the rest of the paper. A complete set of operational definitions of these concepts is available in Refs.\ \cite{ZBIJoT,BZInTech}.\\
\textbf{System}. With the term \textit{system} we mean a set of material particles, of one or more kinds, such that, at each instant of time, the particles of each kind are contained within a given region of space.  Regions of space containing different kinds of particles can overlap and even coincide. If the boundary surfaces of the regions of space which contain the particles of the systems are all \textit{walls},\textit{ i.e.}, surfaces which cannot be crossed by material particles, the system is called \textit{closed}.\\
 \textbf{Property}. Any system is endowed with a set of reproducible measurement procedures; each procedure defines a \textit{property} of the system.\\
 \textbf{State}. The set of all the values of the properties of a system, at a given instant of time, defines the \textit{state} of the system at that instant.\\
\textbf{External force field}. A system can be in contact with other matter, or surrounded by empty space; moreover, force fields due to external matter can act in the region of space occupied by the system. If, at an instant of time, all the particles of the system are removed from the respective regions of space and brought far away, but a force field is still present in the region of space (previously) occupied by the system, then this force field is called an \textit{external force field}. An external force field can be either gravitational, or electric or magnetic, or a superposition of the three.\\
\textbf{Environment of a system}. Consider the union of all the regions of space spanned by a system during its entire time evolution. If no other material particles, except those of the system, are present in the region of space spanned by the system or touches the boundary of this region, and if the external force field in this region is either vanishing or stationary, then we say that the system is \textit{isolated}. Suppose that an isolated system $I$ can be divided into two subsystems, $A$ and $B$. Then, we can say that $B$ is the \textit{environment} of $A$ and viceversa.\\
\textbf{System separable and uncorrelated from its environment}. If, at a given instant of time, two systems $A$ and $B$ are such that the force field produced by $B$ is vanishing in the region of space occupied by $A$ and viceversa, then we say that $A$ and $B$ are \textit{separable} at that instant. The energy of a system $A$ is defined (see Section 3) only for the states of $A$ such that $A$ is separable from its environment. Consider, for instance, the following simple example from mechanics. Let $A$ and $B$ be rigid bodies in deep space, far away from any other object and subjected to a mutual gravitational force. Then, the potential energy of the composite system $AB$ is defined, but that of $A$ and of $B$ is not. For a system A which is separable from its environment, any change in either gravitational, or electric, or magnetic field in the region of space occupied by the system is due to a change of this region, i.e., to a displacement of the system.\\
If, at a given instant of time, two systems $A$ and $B$ are such that the outcomes of the measurements performed on $B$ are statistically independent of those of the measurements performed on $A$, and viceversa, we say that $A$ and $B$ are \textit{uncorrelated from each other} at that instant. The entropy of a system $A$ is defined in this paper only for the states of $A$ such that $A$ is separable and uncorrelated from its environment.\\
\textbf{Process}. We call \textit{process} of a system $A$ from state $A_1$ to state $A_2$ the time evolution of the isolated system $AB$ from $(AB)_1$ (with $A$ in state $A_1$) to $(AB)_2$ (with $A$ in state $A_2$), where $B$ is the environment of $A$.\\
\textbf{Reversible process}. A process of $A$ is \textit{reversible} if the isolated system $AB$ can undergo a time evolution which restores it in its initial state $(AB)_1$. A process of a system $A$ is called a \textit{cycle} for $A$ if the final state $A_2$ coincides with the initial state $A_1$. A cycle for $A$ is not necessarily a cycle for $AB$.\\
\textbf{Weight process}. An \textit{elementary mechanical system} is a system such that the only admissible change of state for it is a space translation in a uniform external force field; an example is a particle which can only change its height in a uniform external gravitational field. A process of a system $A$ from state $A_1$ to $A_2$, such that both in $A_1$ and in $A_2$ system $A$ is separable from its environment, is a \textit{weight process} for $A$ if the only net effect of the process in the environment of $A$ is the change of state of an elementary mechanical system.\\
\textbf{Equilibrium states}. An \textit{equilibrium} state of a system is a state such that the system is separable, the state does not vary with time, and it can be reproduced while the system is isolated. An
equilibrium state of a closed system $A$ in which $A$ is uncorrelated from its environment $B$, is called a \textit{stable equilibrium state} if it cannot be modified by any process between states in which $A$ is separable and uncorrelated from its environment such that neither  the geometrical configuration of the walls which bound the regions of space $\RA{}$ where the constituents of $A$ are contained, nor the state of the environment $B$ of $A$ have net changes. Two systems, $A$ and $B$, are in \textit{mutual stable equilibrium} if the composite system $AB$ (\textit{i.e.}, the union of both systems) is in a stable equilibrium state.\\
\textbf{Weight polygonal and work in a weight polygonal.} Consider an ordered set of \emph{n} states of a closed system $A$, $(A_1, A_2, ... , A_n)$, such that in each of these states $A$ is separable from its environment. If \emph{n} - 1 weight processes exist, which interconnect $A_1$ and $A_2$, ... , $A_{n-1}$ and $A_n$, regardless of the direction of each process, we say that $A_1$ and $A_n$ can be interconnected by a \textit{weight polygonal}.
For instance, if weight processes $A_1\xrightarrow{\rm w}A_2$ and $A_3\xrightarrow{\rm w}A_2$ exist for $A$,
we say that $A_1\xrightarrow{\rm w}A_2\xleftarrow{\rm w}A_3$ is a weight polygonal for $A$ from $A_1$ to $A_3$. We call work done by $A$ in a weight polygonal from $A_1$ to $A_n$ the sum of the works done by $A$ in the weight processes with direction from $A_1$ to $A_n$ and the opposites of the works done by $A$ in the weight processes with direction from $A_n$ to $A_1$ \cite{Zanchini86,Zanchini88}. The work done by $A$ in a weight polygonal from $A_1$ to $A_n$ will be denoted by $W_{1n}^{A\xrightarrow{\rm wp}}$ ; its opposite will be called work received by $A$ in a weight polygonal from $A_1$ to $A_n$ and will be denoted by $W_{1n}^{A\xleftarrow{\rm wp}}$. Clearly, for a given weight polygonal, $W_{1n}^{A\xleftarrow{\rm wp}} = - W_{1n}^{A\xrightarrow{\rm wp}} = W_{n1}^{A\xrightarrow{\rm wp}}$. For the example of weight polygonal $A_1\xrightarrow{\rm w}A_2\xleftarrow{\rm w}A_3$  considered above, we have
\begin{equation}\label{workinpolygonal}
W_{13}^{A\xrightarrow{\rm wp}} = W_{12}^{A\rightarrow} - W_{32}^{A\rightarrow} .
\end{equation}

\section{Postulates and Assumptions}

In this paper, we call \textit{Postulates} the axioms which have a completely general validity and \textit{Assumptions} the additional axioms whose domain of validity could be completely general or not, and identifies the domain of validity of our treatment.
In this section, we list at once all the Postulates and, for ease of reference, also a preview of the Assumptions. However, the assumptions require concepts and notation that we introduce later. Therefore, they will  acquire meaning and will be repeated along our deductive development by introducing them immediately before they become necessary for the subsequent logical development. This approach allows us to emphasize which results require which assumptions.

\begin{postulate}\label{Postulate1} Every pair of states ($A_1$, $A_2$) of a closed system $A$, such that $A$ is separable from its environment in both states, can be interconnected by means of a weight polygonal for $A$.
The works done by a system in any two weight polygonals between the same initial and final states are identical.
\end{postulate}

\begin{comment} In Ref. \cite{Zanchini86} it is proved that, in sets of states where sufficient conditions of interconnectability by weight processes hold, Postulate \ref{Postulate1} can be proved as a consequence of the traditional form of the First Law, which concerns weight processes (or adiabatic processes).
\end{comment}

\begin{postulate}\label{Postulate2} Among all the states of a system $A$ such that the constituents of $A$ are contained in a given set of regions of space $\RA{}$, there is a stable equilibrium state for every value $E^A$ of the energy of
$A$.
\end{postulate}

\begin{postulate}\label{Assumption2} Starting from any state in which the system is separable from its environment, a closed system $A$ can be changed to a stable equilibrium state with the same energy by means of a zero work weight process for $A$ in which the regions of space occupied by the constituents of $A$ have no net changes.
\end{postulate}

\begin{postulate}\label{Postulate3} There exist systems, called normal systems, whose energy has no upper bound. Starting from any state in which the system is separable from its environment, a normal system $A$ can be changed to a non-equilibrium state with arbitrarily higher energy (in which $A$ is separable from its environment) by means of a weight process for $A$ in which the regions of space occupied by the constituents of $A$ have no net changes. \end{postulate}

\begin{comment}
The restriction to normal
closed systems is adopted here in the interest of simplicity, in order to focus the
attention of the reader on the main result of the
paper, namely that of avoiding the use of the concept of thermal reservoir
in the foundations of thermodynamics. The extension of the treatment to \textit{special
systems} and \textit{open systems} will
be presented elsewhere.
\end{comment}

\begin{previewassumption}For
any given pair of states ($A_1$, $A_2$) of any closed system $A$ such that $A$ is separable and uncorrelated from its environment, it is always possible to find or to include in the environment of $A$ a system $B$ which has a stable equilibrium state $B_{\textrm{se}1}$ such
that the states $A_1$ and $A_2$ can be interconnected by means of
a reversible weight process for  $AB$, standard with respect to
$B$, in which system $B$ starts from state $B_{\textrm{se}1}$.
\end{previewassumption}

\begin{previewassumption}The function $f^{B\rightarrow C}_{11}( E^B)$ defined (see Lemma \ref{Lemma3}, below) through the set of pairs of processes $\{(\Pi^{B_{\textrm{se}1}}_{XB\textrm{rev}}; \Pi^{C_{\textrm{se}1}}_{XC\textrm{rev}})\}$ (also defined in Lemma \ref{Lemma3}, below) is differentiable in $E^B_{\textrm{se}1}$; in symbols
\begin{equation*}
\lim_{E^B \rightarrow E^B_{\textrm{se}1}} \frac{f^{B\rightarrow C}_{11}( E^B) - f^{B\rightarrow C}_{11}( E^B_{\textrm{se}1})}{E^B - E^B_{\textrm{se}1}} = \frac{\textrm{d}f^{B\rightarrow C}_{11}}{\textrm{d}E^B}\bigg|_{E^B_{\textrm{se}1}} \;\; .
\end{equation*}
\end{previewassumption}

\begin{previewassumption}For every system $B$ and every choice of the regions of space $\RB{}$ occupied by the constituents of $B$, the temperature of the stable equilibrium states of $B$ (as defined in Section {\ref{TemperatureDef}}, below) is a continuous function of the energy of $B$ and is vanishing only in the stable equilibrium state with the lowest energy for the given regions of space $\RB{}$.
\end{previewassumption}

 In the last section of the paper we extend our operational definition of entropy to a broader class of system models by relaxing Assumption 1 as follows.

\begin{previewassumptionrelaxed}Any given state $A_1$ of any closed system $A$ such that $A$ is separable and uncorrelated from its environment, either belongs to a single set $\Sigma_A$ where every pair of states fulfills Assumption 1 or it can be an intermediate state of at least a composite weight process for $A$ that connects two states of the set $\Sigma_A$.
\end{previewassumptionrelaxed}

\section{Definition of energy for a closed system.}
Let ($A_1$, $A_2$) be any pair of states
of a system $A$, such that $A$ is separable from its environment in both states. We call \emph{energy difference} between
states $A_2$ and $A_1$ the work received by $A$ in any weight polygonal from $A_1$ to $A_2$, expressed as
\begin{equation}\label{energy}
E^A_2 - E^A_1 = - W_{12}^{A\xrightarrow{\rm wp}} = W_{12}^{A\xleftarrow{\rm wp}}.
\end{equation}
The First Law yields the following consequences: \\
(\emph{a}) the energy difference between two states $A_2$ and $A_1$ depends
only on the states $A_1$ and $A_2$;\\
(\emph{b}) (\emph{additivity of energy differences}) consider a pair of states $(AB)_1$
and $(AB)_2$ of a composite system $AB$, and denote by $A_1, B_1$ and $A_2, B_2$ the corresponding states of $A$ and $B$; then, if $A$, $B$ and $AB$ are separable from their environment in the states considered,
\begin{equation}\label{additivity}
E^{AB}_2 - E^{AB}_1 = E^A_2 - E^A_1 + E^B_2 - E^B_1 \;\; ;
\end{equation}
(\emph{c}) (\emph{energy is a property}) let $A_0$ be a reference
state of a system $A$, in which $A$ is separable from its environment, to which we assign an arbitrarily chosen value of energy $E^A_0$; the value of the energy of $A$ in any other state $A_1$ in which $A$ is separable from its environment is determined uniquely by
\begin{equation}\label{energyabs}
E^A_1 = E^A_0 + W_{01}^{A\xleftarrow{\rm wp}} \;\; ,
\end{equation}
where $W_{01}^{A\xleftarrow{\rm wp}}$ is the work received by $A$ in any weight polygonal for $A$ from $A_0$ to $A_1$.\\ Simple proofs of these consequences can be found in Ref. \cite{Zanchini86}, and will not be repeated here.

\begin{comment} The additivity of energy implies that the union of two or more normal systems, each separable from its environment, is a normal system to which Postulate \ref{Postulate3} applies.\\
In traditional treatments of thermodynamics only normal systems are considered, without an
explicit mention of this restriction. Moreover, Postulate \ref{Postulate3} is
\textit{not stated, but it is used}, for example in theorems where
one says that any amount of work can be transferred to a thermal
reservoir by a stirrer. Any system whose constituents have
translational, rotational or vibrational degrees of freedom is a
normal system. On the other hand, quantum theoretical model
systems, such as spins, qubits, qudits, etc., whose energy is
bounded also from above, are special systems.
\end{comment}

\section{\label{TemperatureDef}Definition of temperature of a stable equilibrium state}

\begin{lmm}\label{Lemma1}
\textbf{Uniqueness of the stable equilibrium
state  for a given value of the energy}. There can be no pair of
different stable equilibrium states of a closed system $A$ with
identical regions of space $\RA{}$ and the same value of the
energy $E^A$. \dimostrazione{Lemma}{Lemma1}{Since $A$ is closed and in any stable equilibrium state it is separable and uncorrelated from its environment, if two such states existed, by Postulate \ref{Assumption2} the system could be changed from one to the other by means of a zero-work weight process, with no change of the regions of space occupied by the constituents of $A$ and no change of the state of the environment of $A$. Therefore, neither would satisfy the definition of stable equilibrium state.}
\end{lmm}

\begin{thm}\label{Theorem1}
\textbf{Impossibility of a Perpetual Motion Machine of the Second Kind (PMM2)}. If  a normal system $A$ is in a stable equilibrium state, it is impossible to lower its energy by means of a weight process for $A$ in which the regions of space occupied by the constituents of $A$ have no net change. \dimostrazione{Theorem}{Theorem1}{
Suppose that, starting from a stable equilibrium state $A_{se}$ of
$A$, by means of a weight process $\Pi_1$ with positive work
$W^{A\rightarrow}=W>0$, the energy of $A$ is lowered and the
regions of space $\RA{}$ occupied by the constituents of $A$ have no net change. On account of Postulate \ref{Postulate3}, it would be possible to perform a weight process $\Pi_2$ for $A$ in which its regions of space $\RA{}$  have no net
change, the weight $M$ is restored to its initial state so that
the positive amount of energy $W^{A\leftarrow}=W>0$ is supplied
back to $A$, and the final state of $A$ is a non-equilibrium state,
namely, a state clearly different from $A_{se}$. Thus, the
composite zero-work weight process ($\Pi_1$, $\Pi_2$) would
violate the definition of stable equilibrium state. }
\end{thm}

\begin{comment} \textbf{Kelvin-Planck statement of the Second Law}.
As noted in Refs.\ \cite{HK} and \cite[p.64]{GB}, the impossibility
of a PMM2, which is
also known as the \textit{Kelvin-Planck statement of the Second
Law}, is a corollary of the definition of stable equilibrium
state, provided that we adopt the (usually implicit) restriction
to normal systems.
\end{comment}

\begin{definition} \textbf{Weight process for $AB$, standard with respect to
$B$.} Given a pair of states $(A_1, A_2)$ of a system $A$, such that $A$ is separable from its environment, and a system $B$ in the environment of $A$, we call \textit{weight process for $AB$, standard with
respect to $B$}  a weight process
$A_1B_{\textrm{se}1}\xrightarrow{\rm w} A_2B_{\textrm{se}2}$ for
the composite system $AB$ in which the end states of $A$ are the
given states $A_1$ and $A_2$, and the end states of $B$ are stable
equilibrium states with identical regions of space $\RB{}$.   For
a weight process for $AB$, standard with respect to $B$, we denote
the final energy of system $B$ by the symbol $E^B_{\rm se2}\big|_{A_1 A_2}^{{\rm sw,}B_{\rm se1}}$
or, if the process is reversible, $E^B_{\rm se2rev}\big|_{A_1 A_2}^{{\rm sw,}B_{\rm se1}}$ (when the context allows it, we simply denote them by $E^B_{\rm se2}$ and $E^B_{\rm se2rev}$, respectively).
\end{definition}

\begin{comment} The term \virgolettedue{standard with respect to
$B$} is a shorthand to express the conditions that: 1) the end
states of $B$ are stable equilibrium, and 2) the regions of space
$\RB{\rm se1}$ and $\RB{\rm se2}$ are identical. The regions of
space $\RA{1}$ and $\RA{2}$, instead, need not be identical.
\end{comment}

\begin{assumption}\label{Assumption3} For
any given pair of states ($A_1$, $A_2$) of any closed system $A$ such that $A$ is separable and uncorrelated from its environment, it is always possible to find or to include in the environment of $A$ a system $B$ which has a stable equilibrium state $B_{\textrm{se}1}$ such
that the states $A_1$ and $A_2$ can be interconnected by means of
a reversible weight process for  $AB$, standard with respect to
$B$, in which system $B$ starts from state $B_{\textrm{se}1}$.
\end{assumption}

\begin{comment}
If, for a given pair of states ($A_1$, $A_2$), a stable equilibrium state $B_{\textrm{se}1}$ of $B$ fulfills Assumption \ref{Assumption3}, then any other stable equilibrium state of $B$ with the same regions of space and
with an energy value higher than that of $B_{\textrm{se}1}$
fulfills Assumption \ref{Assumption3}, as well. Therefore, for a given pair of states ($A_1$, $A_2$) of a system $A$ and a selected system
$B$, there exists infinite different choices for
$B_{\textrm{se}1}$.\end{comment}

\begin{comment}
Since Postulates \ref{Postulate1} to \ref{Assumption2} can be considered as having a completely general validity, our new operational definition of entropy for a normal and closed system $A$ applies to any set of states such that $A$ is separable and uncorrelated from its environment, and for which the crucial Assumption \ref{Assumption3} holds for every pair of states in the set.
In the next remark, we show that in quantum statistical mechanics (QSM) Assumption \ref{Assumption3} holds for all the stable equilibrium states and for a large class of nonequilibrium states, for both large and few-particle systems. Moreover, in subsequent remarks we also demonstrate the full compatibility of our construction with the general framework of QSM.
\end{comment}

\begin{comment}\textbf{Plausibility of Assumption \ref{Assumption3} in the framework of quantum statistical mechanics.}
Let us illustrate the range of validity of Assumption \ref{Assumption3} within the framework of quantum statistical mechanics by showing that a broad  (albeit not all-inclusive)  class of pairs ($A_1$,$A_2$)  of nonequilibrium states of any system $A$ can be interconnected by a quantum mechanical unitary process (i.e., a reversible weight process) of a composite system $AB$ such that the initial and final states of a system $B$ are stable equilibrium states.
\\
To this end, given the system $A$, consider a system $B$  with Hilbert space of equal dimensionality ($\dim\Hil_B=\dim\Hil_A$) but different Hamiltonian operator ($H^B\ne H^A$), and consider the states $B_{\textrm{se}1}=\rho^B_1=\exp(-\ln Z_1^B-\beta_1H^B)$ and $B_{\textrm{se}2}=\rho^B_2=\exp(-\ln Z_2^B-\beta_2H^B)$ where $\rho$ denotes the density operator (i.e., the state representative in QSM) and $Z$ the partition function defined by the condition $\Tr\rho=1$.
\\
Next, consider the eigenvalue problems $H^A|\psi^A_{j}\rangle =\varepsilon_{j}^A |\psi^A_{j}\rangle$ and $H^B|\psi^B_{j}\rangle =\varepsilon_{j}^B |\psi^B_{j}\rangle$ and assume that the index $j$ ranks energy eigenvalues in increasing order for both $A$ and $B$.
Finally, consider the states of $A$ constructed as follows $A_1=\rho^A_1=\sum_j\exp(-\ln Z_2^B-\beta_2\varepsilon_{j}^B)|\psi^A_{j}\rangle\langle\psi^A_{j}|$ and $A_2=\rho^A_2=\sum_j\exp(-\ln Z_1^B-\beta_1\varepsilon_{j}^B)|\psi^A_{j}\rangle\langle\psi^A_{j}|$. These states are nonequilibrium states for $A$ because the $\varepsilon_{j}^B$'s are not the eigenvalues of $H^A$ but those of $H^B\ne H^A$. Notice, however,  that they commute with $H^A$ and have decreasing diagonal elements (probabilities), so that $E^A_1=\Tr\rho^A_1 H^A$ is minimal among the states with $-k\Tr\rho\ln\rho=-k\Tr\rho^A_1\ln\rho^A_1=S^A_1$ and $E^A_2=\Tr\rho^A_2 H^A$ is minimal among the states with $-k\Tr\rho\ln\rho=-k\Tr\rho^A_2\ln\rho^A_2=S^A_2$. From these nonequilibrium states we can access, via unitary evolutions of $A$ (i.e., by means of reversible weight processes for $A$), a large number  of nonequilibrium states corresponding to all possible rearrangement of the order of the diagonal elements in the $H^A$ representation.
\\
Now, we need only show that the process $\rho_1^A\otimes\rho_1^B\rightarrow\rho_2^A\otimes\rho_2^B$ can be done by means of a unitary evolution for the composite system $AB$. This is a straightforward consequence of the well-known fact (see, e.g., \cite[p.351]{vonNeumann} and \cite[p.136]{HG}) that by designing a suitable time dependent Hamiltonian that starts and ends with the initial, separable one ($H^{AB}=H^A\otimes I^A+I^B\otimes H^B$) it is possible to change any density operator unitarily into any other one with the same set of eigenvalues. Indeed, the eigenvalues of  $\rho_1^A$,   $\rho_2^B$, $\rho_2^A$, and $\rho_1^B$ are, respectively, $(p_j)_1^A=(p_j)_2^B=\exp(-\ln Z_2^B-\beta_2\varepsilon_{j}^B)$ and $(p_j)_2^A=(p_j)_1^B=\exp(-\ln Z_1^B-\beta_1\varepsilon_{j}^B)$. Therefore, by our particular construction, operators $\rho_1^A\otimes\rho_1^B$ and $\rho_2^A\otimes\rho_2^B$ have the same eigenvalues $(p_{ij})^{AB}_1=(p_i)_1^A(p_j)_1^B=(p_{ij})^{AB}_2=(p_i)_2^A(p_j)_2^B$.
\\
It is noteworthy that if we repeat the above considerations by using a system $B$ identical to $A$, i.e., for  identical Hamiltonian operators ($H^B= H^A$), then states $A_1$ and $A_2$ are stable equilibrium and we confirm that, in the QSM context, Assumption \ref{Assumption3} is always fulfilled for all pairs of stable equilibrium states.
\\
In summary, in a QSM context in which only unitary protocols are conceivable, the domain of validity of Assumption \ref{Assumption3} includes all but only: (1)  the (stable equilibrium) canonical density operators of $A$; and (2) the pairs of nonequilibrium density operators of $A$ each of which can be unitarily reduced to a form with eigenvalues that are canonically distributed with respect to the same but otherwise arbitrary set of energy levels $\varepsilon_{j}^B$ (not necessarily the eigenvalues of $H^A$).
\\
However, if one accepts that, at least in some limiting sense,  non-unitary protocols may be within feasible reach or are at least conceivable, then the domain of validity can be broadened to include all  nonequilibrium density operators. This would be the case if we could smoothly control each eigenvalue of the given $\rho^A_1$ and $\rho^A_2$ so that each of them follows   the path of steepest energy descent \cite[Eq.75]{Beretta08} to the corresponding canonical density operator with entropy $-k\Tr\rho^A_1\ln\rho^A_1$ and $-k\Tr\rho^A_2\ln\rho^A_2$, respectively.
\end{comment}

\begin{thm}\label{Theorem2}
Given a pair of states ($A_1$, $A_2$)
of a system $A$ such that $A$ is separable and uncorrelated from its environment, a system $B$ in the environment of $A$, and an initial stable
equilibrium state $B_{\textrm{se}1}$, among all the weight
processes for $AB$, standard with respect to $B$, in which $A$
goes from $A_1$ to $A_2$ and $B$ begins in state
$B_{\textrm{se}1}$, the energy $E^B_{\rm se2}\big|_{A_1 A_2}^{{\rm sw,}B_{\rm se1}}$ of system $B$ in its final state
has a lower bound, $E^B_{\rm se2rev}\big|_{A_1 A_2}^{{\rm sw,}B_{\rm se1}}$, which is reached if and only if the process is
reversible. Moreover, for all such reversible processes, system
$B$ ends in the same stable equilibrium state
$B_{\textrm{se}2\textrm{rev}}$. \dimostrazione{Theorem}{Theorem2}{Consider a weight process for $AB$,
standard with respect to $B$, $\Pi_{AB}=\Wcomp{A}{B}$, a
reversible weight processes for $AB$, standard with respect to
$B$, $\Pi_{AB\rm rev}=\Wcomprev{A}{B}$, and the corresponding final
energies  of $B$, respectively, $E^B_{\rm se2}\big|_{A_1 A_2}^{{\rm sw,}B_{\rm se1}}$ and $E^B_{\rm se2rev}\big|_{A_1 A_2}^{{\rm sw,}B_{\rm se1}}$. We will prove that:
\\(a) $E^B_{\rm se2rev}\big|_{A_1 A_2}^{{\rm sw,}B_{\rm se1}} \leq E^B_{\rm se2}\big|_{A_1 A_2}^{{\rm sw,}B_{\rm se1}}$;
\\ (b) if also $\Pi_{AB}$ is reversible, then
$E^B_{\rm se2}\big|_{A_1 A_2}^{{\rm sw,}B_{\rm se1}} = E^B_{\rm se2rev}\big|_{A_1 A_2}^{{\rm sw,}B_{\rm se1}}$, and the end stable equilibrium state of $B$ is
the same, \textit{ i.e.}, $B_{\textrm{se}2}=B_{\textrm{se}2\textrm{rev}}$;\\
(c) if $E^B_{\rm se2}\big|_{A_1 A_2}^{{\rm sw,}B_{\rm se1}} = E^B_{\rm se2rev}\big|_{A_1 A_2}^{{\rm sw,}B_{\rm se1}}$, then also $\Pi_{AB}$ is
reversible.
\spazio\textit{Proof of (a)}.  Let us suppose, \textit{ab absurdo},
that  the energy of $B$ in
state $B_{\textrm{se}2}$ is lower than that in state
$B_{\textrm{se}2\textrm{rev}}$. Then, the composite process
($-\Pi_{AB\rm rev}$, $\Pi_{AB}$)
would be a weight process for $B$ in which, starting from the stable equilibrium state $B_{\textrm{se}2\textrm{rev}}$, the energy of $B$ is lowered and
its regions of space  have no net changes, in contrast with
Theorem \ref{Theorem1}. Therefore, $E^B_{\rm se2rev}\big|_{A_1 A_2}^{{\rm sw,}B_{\rm se1}} \leq E^B_{\rm se2}\big|_{A_1 A_2}^{{\rm sw,}B_{\rm se1}}$.

\spazio \textit{Proof of (b)}. If also process $\Pi_{AB}$ is
reversible, then, in addition to $E^B_{\rm se2rev}\big|_{A_1 A_2}^{{\rm sw,}B_{\rm se1}} \leq E^B_{\rm se2}\big|_{A_1 A_2}^{{\rm sw,}B_{\rm se1}}$,
also the relation $E^B_{\rm se2}\big|_{A_1 A_2}^{{\rm sw,}B_{\rm se1}} \leq E^B_{\rm se2rev}\big|_{A_1 A_2}^{{\rm sw,}B_{\rm se1}}$ must hold by
virtue of the proof of a) just given and, therefore, $E^B_{\rm se2rev}\big|_{A_1 A_2}^{{\rm sw,}B_{\rm se1}} = E^B_{\rm se2}\big|_{A_1 A_2}^{{\rm sw,}B_{\rm se1}}$. On account of Postulate \ref{Postulate2} and Lemma \ref{Lemma1}, the final
value of the energy of $B$ determines a unique final stable
equilibrium state of $B$; therefore
$B_{\textrm{se}2}=B_{\textrm{se}2\textrm{rev}}$.

\spazio \textit{Proof of (c)}. Let $\Pi_{AB}$ be  such that
$E^B_{\rm se2}\big|_{A_1 A_2}^{{\rm sw,}B_{\rm se1}} = E^B_{\rm se2rev}\big|_{A_1 A_2}^{{\rm sw,}B_{\rm se1}}$.  Then, the final states
$B_{\textrm{se}2}$ and $B_{\textrm{se}2\textrm{rev}}$ have the
same energy and, being stable equilibrium states, by Lemma \ref{Lemma1} they
must coincide. Thus, the composite process ($\Pi_{AB}$,
$-\Pi_{AB\rm rev}$) is a cycle for the isolated system $ABC$,
where $C$ is the environment of $AB$, where the only effect is the
return of the weight to its initial position. As a consequence,
being a part of a cycle of the isolated system $ABC$, process
$\Pi_{AB}$ is reversible.}
\end{thm}

\begin{thm}\label{Theorem3}
Consider a pair of states ($A_1$, $A_2$)
of a system $A$ such that $A$ is separable and uncorrelated from its environment, and two systems in the environment of $A$, $B$ and $C$, in given initial stable
equilibrium states $B_{\textrm{se}1}$ and $C_{\textrm{se}1}$. Let $\Pi_{AB\textrm{rev}}$ and $\Pi_{AC\textrm{rev}}$ be reversible weight processes
for $AB$ and for $AC$, both from $A_1$ to $A_2$ and standard with respect to $B$ and $C$ respectively, with the given initial states $B_{\textrm{se}1}$ and $C_{\textrm{se}1}$ of $B$ and $C$; let $E^B$ denote, for shorthand, the final energy  $E^B_{\rm se2rev}\big|_{A_1 A_2}^{{\rm sw,}B_{\rm se1}}$ of $B$ in process $\Pi_{AB\textrm{rev}}$ and let $E^C$ denote the final energy  $E^C_{\rm se2rev}\big|_{A_1 A_2}^{{\rm sw,}C_{\rm se1}}$ of $C$ in process $\Pi_{AC\textrm{rev}}$. Then, if $E^B-E^B_{\textrm{se}1}$ is vanishing, $E^C-E^C_{\textrm{se}1}$ is vanishing as well; if $E^B-E^B_{\textrm{se}1}$ is non vanishing, $E^C-E^C_{\textrm{se}1}$ is non vanishing and the ratio $(E^B-E^B_{\textrm{se}1}) \big / (E^C-E^C_{\textrm{se}1})$ is positive. \dimostrazione{Theorem}{Theorem3}{Assume that $E^B-E^B_{\textrm{se}1}$ is vanishing and that $E^C-E^C_{\textrm{se}1}$ is positive, and consider the composite process
($\Pi_{AB\textrm{rev}}, -\Pi_{AC\textrm{rev}}$);
this would be a reversible weight process for $C$ in which the energy change of $C$ is negative, the regions of space occupied by $C$ did not change and the initial state of $C$ is a stable equilibrium state, in contrast with Theorem \ref{Theorem1}. Assume now that $E^B-E^B_{\textrm{se}1}$ is vanishing and that $E^C-E^C_{\textrm{se}1}$ is negative, and consider the composite process ($\Pi_{AC\textrm{rev}},  -\Pi_{AB\textrm{rev}}$); this would be a reversible weight process for $C$ in which the energy change of $C$ is negative, the regions of space occupied by $C$ do not change and the initial state of $C$ is a stable equilibrium state, in contrast with Theorem \ref{Theorem1}. Then, if $E^B-E^B_{\textrm{se}1}$ is vanishing, $E^C-E^C_{\textrm{se}1}$ is vanishing as well.\\
Assume now that the energy change of $B$
is negative, \textit{i.e.}, $E^B-E^B_{\textrm{se}1}
<0$. Clearly, the energy change of $C$ cannot be
zero, because this would imply $E^B-E^B_{\textrm{se}1}=0$. Suppose that the energy change of $C$ is positive, $E^C-E^C_{\textrm{se}1}>0$, and consider the composite process ($\Pi_{AB\textrm{rev}}, -\Pi_{AC\textrm{rev}}$). In this process, which is a cycle for $A$, system $BC$ would have performed a positive work, given (energy balance for $BC$) by the sum of two positive addenda, namely $W = -(E^B-E^B_{\textrm{se}1}) +(E^C-E^C_{\textrm{se}1})$. On account of Postulates \ref{Postulate3} and \ref{Assumption2}, one could supply back to system $C$ a positive work amount equal to $(E^C-E^C_{\textrm{se}1})$ and restore $C$ to its initial state $C_{\rm se1}$
by means of a composite weight process $\Pi_C=C_{\rm
se2}\xrightarrow{\rm w}C_{\rm 3}\xrightarrow{\rm w}C_{\rm se1}$
where $C_3$ has energy $E^C_3=E^C_{\rm se1}$. Thus, the composite
process $(\Pi_{AB\textrm{rev}}, -\Pi_{AC\textrm{rev}}, \Pi_C)$ would be a again a weight
process for $B$ which violates Theorem \ref{Theorem1}. Therefore, if $E^B-E^B_{\textrm{se}1}$ is negative, $E^C-E^C_{\textrm{se}1}$ is negative as well.\\
Let us assume now that, in process $\Pi_{AB\rm rev}$, the energy change of $B$ is
positive. Then, in the reverse process $- \Pi_{AB\rm rev}$, the
energy change of $B$ is negative and, as we have just proved, the
energy change of $C$ in the reverse process - $\Pi_{AC\rm rev}$ must be negative as well. Therefore, in process $\Pi_{AC\rm rev}$, the energy change of $C$ is positive.}
\end{thm}

\begin{lmm}\label{Lemma3}
Consider a pair of systems, $B$ and $C$, a pair of stable equilibrium states of these systems, $B_{\textrm{se}1}$ and $C_{\textrm{se}1}$, and a system $X$ in the environment of $BC$ with an initial state $X_1$ such that: every stable equilibrium state of $B$ with the same regions of space as $B_{\textrm{se}1}$ can be interconnected with $B_{\textrm{se}1}$ by a reversible weight process for $XB$ starting from $(X_1,B_{\textrm{se}1})$; every stable equilibrium state of $C$ with the same regions of space as $C_{\textrm{se}1}$ can be interconnected with $C_{\textrm{se}1}$ by a reversible weight process for $XC$ starting from $(X_1,C_{\textrm{se}1})$.\\ Denote by $\{(\Pi^{B_{\textrm{se}1}}_{XB\textrm{rev}}; \Pi^{C_{\textrm{se}1}}_{XC\textrm{rev}})\}$ the set of all the pairs of reversible weight processes for $XB$ and for $XC$, standard with respect to $B$ and $C$ and with initial states $(X_1, B_{\textrm{se}1})$ and $(X_1, C_{\textrm{se}1})$ respectively, and such that for each pair of processes, the final state $X_2$ of $X$ is the same. Let $\{(B_{\textrm{se}2};C_{\textrm{se}2})\}$ the set of pairs  of final states of $B$ and $C$ which one obtains by the set of pairs of processes $\{(\Pi^{B_{\textrm{se}1}}_{XB\textrm{rev}}; \Pi^{C_{\textrm{se}1}}_{XC\textrm{rev}})\}$, and let $\{(E^B_{\textrm{se}2};E^C_{\textrm{se}2})\}$ be the corresponding values of the energy of $B$ and $C$. Then, the set of pairs of processes $\{(\Pi^{B_{\textrm{se}1}}_{XB\textrm{rev}}; \Pi^{C_{\textrm{se}1}}_{XC\textrm{rev}})\}$ determines a single valued and invertible function from the set $\{E^B_{\textrm{se}2}\}$ to the set $\{E^C_{\textrm{se}2}\}$,
\begin{equation}\label{function}
E^C = f^{B\rightarrow C}_{11}( E^B)\;\; ,
\end{equation}
which is independent on the choice of system $X$ and on the initial state $X_1$ used to construct the set of processes $\{(\Pi^{B_{\textrm{se}1}}_{XB\textrm{rev}}; \Pi^{C_{\textrm{se}1}}_{XC\textrm{rev}})\}$. \dimostrazione{Lemma}{Lemma3}{Choose a system $X$ and an initial state $X_1$ of $X$, and consider a pair of reversible weight processes $(\Pi^{B_{\textrm{se}1}}_{XB\textrm{rev}}; \Pi^{C_{\textrm{se}1}}_{XC\textrm{rev}})$, which belongs to the set $\{(\Pi^{B_{\textrm{se}1}}_{XB\textrm{rev}}; \Pi^{C_{\textrm{se}1}}_{XC\textrm{rev}})\}$. Let $X_2$, $B_{\textrm{se}2}$ and $C_{\textrm{se}2}$ be the final states of $X$, $B$ and $C$ for this pair of processes. Choose now a system $X'$ and an initial state $X'_1$ of $X'$, and consider a pair of reversible weight processes $(\Pi^{B_{\textrm{se}1}}_{X'B\textrm{rev}}; \Pi^{C_{\textrm{se}1}}_{X'C\textrm{rev}})$, which belongs to the set $\{(\Pi^{B_{\textrm{se}1}}_{X'B\textrm{rev}}; \Pi^{C_{\textrm{se}1}}_{X'C\textrm{rev}})\}$. Let $X'_2$, $B_{\textrm{se}3}$ and $C_{\textrm{se}3}$ be the final states of $X$, $B$ and $C$ for this pair of processes. We will prove that, if $B_{\textrm{se}3}$ coincides with $B_{\textrm{se}2}$, then also $C_{\textrm{se}3}$ coincides with $C_{\textrm{se}2}$, so that the correspondence between the final stable equilibrium states of $B$ and $C$ is not affected by either the choice of the auxiliary system, $X$ or $X'$, or the choice of the initial state of the auxiliary system.\\
Consider the composite system $XX'BC$, in the initial state $X_1X'_2B_{\textrm{se}1}C_{\textrm{se}2}$, and consider the composite process $\Pi=(\Pi^{B_{\textrm{se}1}}_{XB\textrm{rev}}, - \Pi^{C_{\textrm{se}1}}_{XC\textrm{rev}}, - \Pi^{B_{\textrm{se}1}}_{X'B\textrm{rev}},  \Pi^{C_{\textrm{se}1}}_{X'C\textrm{rev}})$, where $- \Pi^{C_{\textrm{se}1}}_{XC\textrm{rev}}$ is the reverse of $\Pi^{C_{\textrm{se}1}}_{XC\textrm{rev}}$ and $- \Pi^{B_{\textrm{se}1}}_{X'B\textrm{rev}}$ is the reverse of $\Pi^{B_{\textrm{se}1}}_{X'B\textrm{rev}}$. As easily verified,  $\Pi=(\Pi_{XB\textrm{rev}}, - \Pi_{XC\textrm{rev}}, - \Pi_{X'B\textrm{rev}},  \Pi_{X'C\textrm{rev}})=X_1X'_2B_{\textrm{se}1}C_{\textrm{se}2} \xrightarrow{\rm wrev} X_2X'_2B_{\textrm{se}2}C_{\textrm{se}2} \xrightarrow{\rm wrev} X_1X'_2B_{\textrm{se}2}C_{\textrm{se}1}
\xrightarrow{\rm wrev} X_1X'_1B_{\textrm{se}1}C_{\textrm{se}1}
\xrightarrow{\rm wrev} X_1X'_2B_{\textrm{se}1}C_{\textrm{se}3}$, therefore, the final state of the composite system $XX'BC$, after process $\Pi$, is $X_1 X'_2 B_{\textrm{se}1}C_{\textrm{se}3}$. Therefore, $\Pi$ is a reversible weight process for $C$ in which the regions of space occupied by the constituents of $C$ have no net change. If the energy of $C$ in state $C_{\textrm{se}3}$ were lower than that in the initial state $C_{\textrm{se}2}$, then $\Pi$ would violate Theorem \ref{Theorem1}. If the energy of $C$ in state $C_{\textrm{se}3}$ were higher than that in the initial state $C_{\textrm{se}2}$, then the reverse of $\Pi$ would violate Theorem \ref{Theorem1}. Therefore, the energy of $C$ in state $C_{\textrm{se}3}$ must coincide with the energy of $C$ in state $C_{\textrm{se}2}$, \textit{i.e.}, on account of Postulate \ref{Postulate2} and Lemma \ref{Lemma1}, the state $C_{\textrm{se}3}$ must coincide with $C_{\textrm{se}2}$.}
\end{lmm}

\begin{comment}Lemma \ref{Lemma3} plays a key role in the new definition of temperature that we propose below. On the other hand, the statement and the proof of Lemma \ref{Lemma3} require a considerable effort to be fully understood. Therefore, to help the reader, we show here that the existence of the function $f^{B\rightarrow C}_{11}( E^B)$, as well as its differentiability in $E^B_{\textrm{se}1}$, that we state later in Assumption \ref{Assumption4}, can be easily shown in the usual framework of thermodynamics \textit{a posteriori}, i.e., once temperature and entropy are defined.
Consider systems $B$ and $C$ in their initial stable equilibrium states $B_{\textrm{se}1}$ and $C_{\textrm{se}1}$, with energy values $E^B_{\textrm{se}1}$  and $E^C_{\textrm{se}1}$  (and corresponding temperatures $T^B_{\textrm{se}1}$  and $T^C_{\textrm{se}1}$) and pairs of reversible weight processes for $XB$ and for $XC$, standard with respect to $B$ and $C$, with initial state $X_1$ of the auxiliary system $X$ and such that for each pair of processes, the final state $X_2$ of $X$ is the same, so that the entropy change of $X$ is the same.  In each pair of such processes, the entropy change $\Delta S$ of $C$ is equal to that of $B$, because both are equal to the opposite of the entropy change of $X$. Moreover, the end states of $B$ and $C$ are stable equilibrium states and the regions of space occupied by the systems do not change. For a closed system that occupies a fixed region of space, each stable equilibrium state is determined uniquely either by the value of the energy, or by that of the entropy, since entropy is a strictly increasing function of the energy. Therefore, the final energy values $E^B$ and $E^C$  are determined uniquely by $\Delta S$, so that the latter determines a one-to-one correspondence between the final energy values $E^B$ and $E^C$, that we denote by $E^C=f^{B\rightarrow C}_{11}( E^B)$.  In addition, when $\Delta S$  is very small, also the energy changes $\Delta E^B$  and $\Delta E^C$ are very small, and one has $\Delta E^B\approx T^B_{\textrm{se}1}\Delta S  $, $\Delta E^C\approx T^C_{\textrm{se}1}\Delta S$, so that $\Delta E^C\approx  T^C_{\textrm{se}1}\Delta E^B/T^B_{\textrm{se}1} $, and by considering the limit for $\Delta S\rightarrow 0$ one obtains
\begin{equation}
\lim_{\Delta S\rightarrow 0} \frac{\Delta E^C(\Delta S)}{\Delta E^B(\Delta S)}=\lim_{\Delta E^B\rightarrow 0} \frac{\Delta E^C(\Delta E^B)}{\Delta E^B}=\frac{T^C_{\textrm{se}1}}{T^B_{\textrm{se}1}} \; .
\end{equation}
In another remark, reported after our definition of temperature, we  show, as an example, how the function $f^{B\rightarrow C}_{11}( E^B)$ is construed in the framework of quantum statistical mechanics (Eq.\ \ref{QSMfunction}).
\end{comment}

\begin{lmm}\label{LemmadE}
 For a given pair of systems, $B$ and $C$, consider an arbitrary pair of stable equilibrium states $(B_{\textrm{se}1}, C_{\textrm{se}1})$ % that fulfill Assumption \ref{Assumption3}
 and the set of processes which defines the function  $E^C=f^{B\rightarrow C}_{11}(E^B)$ according to Lemma \ref{Lemma3}. Select another arbitrary stable equilibrium state  $B_{\textrm{se}2}$ of system $B$ and let $C_{\textrm{se}2}$ be the stable equilibrium state of system $C$ such that $E^C_{\textrm{se}2} = f^{B\rightarrow C}_{11}(E^B_{\textrm{se}2})$. Denote by $E^C=f^{B\rightarrow C}_{22}(E^B)$ the function defined by the set of reversible processes $\{(\Pi^{B_{\textrm{se}2}}_{XB\textrm{rev}}; \Pi^{C_{\textrm{se}2}}_{XC\textrm{rev}})\}$ according to Lemma \ref{Lemma3}. Then we have the identity
   \begin{equation}\label{new-equality44}
f^{B\rightarrow C}_{11}(E^B) =  f^{B\rightarrow C}_{22}(E^B) \mbox{ for every } E^B \ .
\end{equation}
 \dimostrazione{Lemma}{LemmadE}{Consider an arbitrary stable equilibrium state  of system $B$ with energy $E^B$ and denote it by $B_{\textrm{se}3}$, i.e. $E^B_{\textrm{se}3}=E^B$, and let $C_{\textrm{se}3}$ be the stable equilibrium state of system $C$ such that $E^C_{\textrm{se}3} = f^{B\rightarrow C}_{22}(E^B)$. Then, the pair of composite processes
$( X_1B_{\textrm{se}1} \xrightarrow{\rm wrev} X_2B_{\textrm{se}2} \xrightarrow{\rm wrev} X_3B_{\textrm{se}3} \;, \; X_1C_{\textrm{se}1} \xrightarrow{\rm wrev} X_2C_{\textrm{se}2} \xrightarrow{\rm wrev} X_3C_{\textrm{se}3})$ exists because $E^C_{\textrm{se}2} = f^{B\rightarrow C}_{11}(E^B_{\textrm{se}2})$ and $E^C_{\textrm{se}3} = f^{B\rightarrow C}_{22}(E^B_{\textrm{se}3})$, and it clearly belongs to the set of pairs of processes which defines the function $E^C=f^{B\rightarrow C}_{11}(E^B)$, therefore, $E^C_{\textrm{se}3} = f^{B\rightarrow C}_{11}(E^B_{\textrm{se}3})$.}
\end{lmm}

\begin{coroll}\label{Corollaryb}
The function $f^{B\rightarrow C}_{11}( E^B)$ defined through the set of pairs of processes $\{(\Pi^{B_{\textrm{se}1}}_{XB\textrm{rev}}; \Pi^{C_{\textrm{se}1}}_{XC\textrm{rev}})\}$ is strictly increasing. \dimostrazione{Corollary}{Corollaryb}{Consider the pairs of stable equilibrium states $(B_{\textrm{se}2}, C_{\textrm{se}2})$ and $(B_{\textrm{se}3}, C_{\textrm{se}3})$, such that $E^C_{\textrm{se}2} = f^{B\rightarrow C}_{11}(E^B_{\textrm{se}2})$, $E^C_{\textrm{se}3} = f^{B\rightarrow C}_{11}(E^B_{\textrm{se}3})$, and $E^B_{\textrm{se}3} > E^B_{\textrm{se}2}$. We will prove that $E^C_{\textrm{se}3} > E^C_{\textrm{se}2}$, \textit{i.e.}, $f^{B\rightarrow C}_{11}(E^B_{\textrm{se}3}) > f^{B\rightarrow C}_{11}(E^B_{\textrm{se}2})$.\\
Consider the pair of composite processes $( X_2B_{\textrm{se}2} \xrightarrow{\rm wrev} X_1B_{\textrm{se}1} \xrightarrow{\rm wrev} X_3B_{\textrm{se}3} \;, \; X_2C_{\textrm{se}2} \xrightarrow{\rm wrev} X_1C_{\textrm{se}1} \xrightarrow{\rm wrev} X_3C_{\textrm{se}3})$, which exists because $E^C_{\textrm{se}2} = f^{B\rightarrow C}_{11}(E^B_{\textrm{se}2})$ and $E^C_{\textrm{se}3} = f^{B\rightarrow C}_{11}(E^B_{\textrm{se}3})$. In this pair of processes, the energy change of $B$, $E^B_{\textrm{se}3} - E^B_{\textrm{se}2}$, is positive. On account of Theorem \ref{Theorem3}, also the energy change of $C$ must be positive, i.e., $E^C_{\textrm{se}3} > E^C_{\textrm{se}2}$.}
\end{coroll}

\begin{comment} Since the function $f^{B\rightarrow C}_{11}( E^B)$ is strictly increasing, it is invertible. The inverse of the function
\begin{equation}\label{function-direct}
E^C = f^{B\rightarrow C}_{11}( E^B)\;\; ,
\end{equation}
will be denoted by
\begin{equation}\label{inverse-function}
E^B = f^{C\rightarrow B}_{11}( E^C)\;\; .
\end{equation}
The domain of function $f^{B\rightarrow C}_{11}$ is the set of all the energy values of system $B$ compatible with the regions of space occupied by the constituents of $B$ in state $B_{\textrm{se}1}$. The domain of function $f^{C\rightarrow B}_{11}$ is the set of all the energy values of system $C$ compatible with the regions of space occupied by the constituents of $C$ in state $C_{\textrm{se}1}$.\end{comment}

\begin{lmm}\label{Lemma4}
Consider three systems $B$, $C$, and $R$,  three stable equilibrium states $B_{\textrm{se}1}$, $C_{\textrm{se}1}$, and  $R_{\textrm{se}1}$, and the functions $E^R = f^{B\rightarrow R}_{11}( E^B)$, $E^C = f^{R\rightarrow C}_{11}( E^R)$, and $E^C = f^{B\rightarrow C}_{11}( E^B)$ defined as in Lemma \ref{Lemma3}. Then,
\begin{equation}\label{function-function}
f^{B\rightarrow C}_{11}( E^B)=f^{R\rightarrow C}_{11}( f^{B\rightarrow R}_{11}( E^B))\;\; .
\end{equation}
 \dimostrazione{Lemma}{Lemma4}{Consider an auxiliary system $X$, the pair of states $(X_1,X_2)$, and the three processes  $\Pi^{B_{\textrm{se}1}}_{XB\textrm{rev}}$, $\Pi^{C_{\textrm{se}1}}_{XC\textrm{rev}}$, $\Pi^{R_{\textrm{se}1}}_{XR\textrm{rev}}$, respectively defined as follows: $\Pi^{B_{\textrm{se}1}}_{XB\textrm{rev}}$ is a reversible weight process for $XB$ with initial and final states $X_1$ and $X_2$ for $X$, and initial state $B_{\textrm{se}1}$ for $B$;  $\Pi^{C_{\textrm{se}1}}_{XC\textrm{rev}}$ is a reversible weight process for $XC$ with initial and final states $X_1$ and $X_2$ for $X$, and initial state $C_{\textrm{se}1}$ for $C$; $\Pi^{R_{\textrm{se}1}}_{XR\textrm{rev}}$ is a reversible weight process for $XR$ with initial and final states $X_1$ and $X_2$ for $X$, and initial state $R_{\textrm{se}1}$ for $R$. Let us denote by $E^B_{\textrm{se}2}$, $E^C_{\textrm{se}2}$, $E^R_{\textrm{se}2}$ the energy of the final states of $B$, $C$ and $R$, respectively.
The pair of processes $(\Pi^{B_{\textrm{se}1}}_{XB\textrm{rev}},\Pi^{R_{\textrm{se}1}}_{XR\textrm{rev}})$ belongs to the set of processes $\{(\Pi^{B_{\textrm{se}1}}_{XB\textrm{rev}},\Pi^{R_{\textrm{se}1}}_{XR\textrm{rev}})\}$ that defines according to Lemma \ref{Lemma3} the function $E^R = f^{B\rightarrow R}_{11}( E^B)$, therefore,
\begin{equation}\label{ffBR}
E^R_{\textrm{se}2} = f^{B\rightarrow R}_{11}( E^B_{\textrm{se}2}) \ .
\end{equation}
The pair of processes $(\Pi^{R_{\textrm{se}1}}_{XR\textrm{rev}},\Pi^{C_{\textrm{se}1}}_{XC\textrm{rev}})$ belongs to the set of processes $\{(\Pi^{R_{\textrm{se}1}}_{XR\textrm{rev}},\Pi^{C_{\textrm{se}1}}_{XC\textrm{rev}})\}$ that defines according to Lemma \ref{Lemma3} the function $E^C = f^{R\rightarrow C}_{11}( E^R)$, therefore,
\begin{equation}\label{ffRC}
E^C_{\textrm{se}2} = f^{R\rightarrow C}_{11}( E^R_{\textrm{se}2}) \ .
\end{equation}
The pair of processes $(\Pi^{B_{\textrm{se}1}}_{XB\textrm{rev}},\Pi^{C_{\textrm{se}1}}_{XC\textrm{rev}})$ belongs to the set of processes $\{(\Pi^{B_{\textrm{se}1}}_{XB\textrm{rev}},\Pi^{C_{\textrm{se}1}}_{XC\textrm{rev}})\}$ that defines according to Lemma \ref{Lemma3} the function $E^C = f^{B\rightarrow C}_{11}( E^B)$, therefore,
\begin{equation}\label{ffBC}
E^C_{\textrm{se}2} = f^{B\rightarrow C}_{11}( E^B_{\textrm{se}2}) \ .
\end{equation}
From (\ref{ffBR}) and (\ref{ffRC}) it follows that
\begin{equation}\label{fff}
E^C_{\textrm{se}2} =f^{R\rightarrow C}_{11}( f^{B\rightarrow R}_{11}( E^B_{\textrm{se}2}))\;\; .
\end{equation}
Comparing (\ref{fff}) and (\ref{ffBC}) we find
\begin{equation}\label{fff2}
f^{B\rightarrow C}_{11}( E^B_{\textrm{se}2}) =f^{R\rightarrow C}_{11}( f^{B\rightarrow R}_{11}( E^B_{\textrm{se}2}))\;\; .
\end{equation}
Equation (\ref{function-function}) follows immediately from (\ref{fff2}) by repeating the above for all possible choices of the pair of states $(X_1,X_2)$.}
\end{lmm}

\begin{assumption}\label{Assumption4} The function $f^{B\rightarrow C}_{11}( E^B)$ defined through the set of pairs of processes $\{(\Pi^{B_{\textrm{se}1}}_{XB\textrm{rev}}; \Pi^{C_{\textrm{se}1}}_{XC\textrm{rev}})\}$ is differentiable in $E^B_{\textrm{se}1}$; in symbols
\begin{equation}\label{limit}
\lim_{E^B \rightarrow E^B_{\textrm{se}1}} \frac{f^{B\rightarrow C}_{11}( E^B) - f^{B\rightarrow C}_{11}( E^B_{\textrm{se}1})}{E^B - E^B_{\textrm{se}1}} = \frac{\textrm{d}f^{B\rightarrow C}_{11}}{\textrm{d}E^B}\bigg|_{E^B_{\textrm{se}1}} \;\; .
\end{equation}
Below, in the Remark that follows the definition of temperature, we show that Assumption \ref{Assumption4} is fulfilled within the quantum statistical mechanics, at least within the range of validity of Assumption \ref{Assumption3}.
\end{assumption}

\begin{coroll}\label{Corollaryc}
The inverse function $E^B = f^{C\rightarrow B}_{11}( E^C)$ is differentiable in
$E^C_{\textrm{se}1}$, moreover if  $\textrm{d}f^{B\rightarrow C}_{11}/\textrm{d}E^B\big|_{E^B_{\textrm{se}1}}\ne 0$ then
\begin{equation}\label{derivative-inverse}
\frac{\textrm{d}f^{C\rightarrow B}_{11}}{\textrm{d}E^C}\bigg|_{E^C_{\textrm{se}1}} = \frac{1}{\displaystyle\frac{\textrm{d}f^{B\rightarrow C}_{11}}{\textrm{d}E^B}\bigg|_{E^B_{\textrm{se}1}}}\;\; .
\end{equation}
\vskip-3mm\noindent
 \dimostrazione{Corollary}{Corollaryc}{Since Assumption \ref{Assumption4} holds for any pair of systems, by exchanging $B$ with $C$ it implies that also the function $f^{C\rightarrow B}_{11}( E^C)$ is differentiable. Equation (\ref{derivative-inverse}) follows from the theorem on the derivative of the inverse function.}
\end{coroll}

\begin{definition}
\textbf{Temperature of a stable equilibrium state.}
Let $R$ be a \textit{reference system}, and let $R_{\rm se1}$ be a
\textit{reference stable equilibrium state} of $R$.
Both $R$ and $R_{\rm se1}$ are fixed once and for all, and a positive real number, $T^R_{\rm se1}$, chosen arbitrarily, is associated with
$R_{\rm se1}$ and called \textit{temperature} of $R_{\rm se1}$. Let $B$ be any system, and $B_{\textrm{se}1}$ any stable equilibrium state of $B$.\\
Let us consider the set of pairs of processes $\{(\Pi^{R_{\textrm{se}1}}_{XR\textrm{rev}}; \Pi^{B_{\textrm{se}1}}_{XB\textrm{rev}})\}$, where $\Pi^{R_{\textrm{se}1}}_{XR\textrm{rev}}$ is any reversible weight processes for $XR$ standard with respect $R$ and with initial state $R_{\textrm{se}1}$, $\Pi^{B_{\textrm{se}1}}_{XB\textrm{rev}}$ is any reversible weight processes for $XB$ standard with respect $B$ and with initial state $B_{\textrm{se}1}$, and $X$ is a system which can be chosen and changed arbitrarily, as well as the initial state of $X$. On account of Lemma \ref{Lemma3} and of Assumption \ref{Assumption4}, the set of pairs of processes $\{(\Pi^{R_{\textrm{se}1}}_{XR\textrm{rev}}; \Pi^{B_{\textrm{se}1}}_{XB\textrm{rev}})\}$ defines a single valued and invertible function $f^{R\rightarrow B}_{11}( E^R)$, from the energy values of the stable equilibrium states of $R$ with the same regions of space  as $R_{\textrm{se}1}$ to the energy values of the stable equilibrium states of $B$ with the same regions of space  as $B_{\textrm{se}1}$, which is differentiable in $E^R_{\textrm{se}1}$. We define as \textit{temperature} of system $B$ in the stable equilibrium state $B_{\rm se1}$ the quantity
\begin{equation}\label{temperature}
\!\frac{T^B_{\rm se1}}{T^R_{\rm se1}} \!=\!\!\! \lim_{E^R \rightarrow E^R_{\textrm{se}1}} \!\!\!\! \frac{f^{R\rightarrow B}_{11}\!( E^R) - f^{R\rightarrow B}_{11}\!( E^R_{\textrm{se}1})}{E^R - E^R_{\textrm{se}1}} \!=\!
\frac{\textrm{d}f^{R\rightarrow B}_{11}}{\textrm{d}E^R}\bigg|_{E^R_{\textrm{se}1}} \! .
\end{equation}
On account of Corollary \ref{Corollaryb}, $T^B_{\rm se1}$ is non-negative. Since $R$ and $R_{\textrm{se}1}$ have been fixed once and for all, the temperature is a property of $B$, defined for all the stable equilibrium states of $B$. Clearly, the property temperature is defined by Eq.\ (\ref{temperature}) only with respect to the chosen reference state $R_{\rm se1}$ of the reference system $R$ and up to the arbitrary multiplicative constant $T^R_{\rm se1}$.
\end{definition}

\begin{comment}
\textbf{Agreement with quantum statistical mechanics.}
 As an important example, we show that the existence of the single valued function    defined in Lemma \ref{Lemma3}, the differentiability of that function (Assumption \ref{Assumption4}) and the definition of thermodynamic temperature given by Eq. \ref{temperature} are in agreement with the standard description of stable equilibrium states in quantum statistical mechanics (QSM).
\\
In QSM, the Hamiltonian operator $H$ of a system defines the canonical  density operators that represent the stable equilibrium (maximum von Neumann entropy) states, namely, the one-parameter family $\rho_{\rm se}(\beta)=\exp(-\beta\,H)/Z(\beta)$, where $Z(\beta)= \Tr\exp(-\beta\,H)$ is the partition function.  As is well known (see, e.g., \cite[p.392]{vonNeumann} and \cite[p.45]{Katz}), $\rho(\beta)$ is the maximizer of  the von Neumann entropy functional $S_{\rm vN}=-\Boltz\Tr\rho\ln\rho$ subject to the constraints $ \Tr\rho =1$ and $ \Tr\rho H=E$,  where $\beta$ is the Lagrange multiplier associated with latter constraint and is clearly a continuous function of $E$. The relations $ E_{\rm se}(\beta)=-\partial  \ln Z(\beta)/\partial \beta$, $S_{\rm vN,se}(\beta)=\Boltz\beta  E_{\rm se}(\beta) +\Boltz\ln Z(\beta)$, $C(\beta)=-\Boltz\beta^2\partial E_{\rm se}(\beta)/\partial \beta=\Boltz^2\beta^2\partial^2 \ln Z(\beta)/\partial \beta^2=-\beta\partial S_{\rm vN,se}(\beta)/\partial \beta$ determine the mean energy $E$, the von Neumann entropy $S_{\rm vN,se}$ and the heat capacity $C$, respectively, where $\Boltz$ is Boltzmann's constant. Since  $C(\beta)/\Boltz\beta^2$ is positive, relation $E=E_{\rm se}(\beta)$ is monotonic and hence invertible,  to yield $\beta=\beta(E)$. Therefore, we may write $S_{\rm vN,se}(E)=\Boltz E\,\beta(E) +\Boltz\ln Z(\beta(E))$ and $dS_{\rm vN,se}(E)/dE=\Boltz\beta(E)$.
\\
Next, consider  the  function of $E$ defined by the difference between the von Neumann entropies $S_{\rm vN,se}(E)$ and $S_{\rm vN,se}(E_1)$ of the two stable equilibrium states with energies $E$ and  $E_1$, respectively. Assuming that $E_1$ is a fixed value, let us denote this function by \begin{equation}h_1(E)=S_{\rm vN,se}(E)-S_{\rm vN,se}(E_1)\ ,\end{equation} its derivative by $h'_1$, and its inverse by $\ih_1$, so that the inverse function theorem yields the identity
 \begin{equation}\ih'_1(h_1(E))= \frac{1}{ h'_1 (\ih_1(h_1(E))) }  \ .\label{inverseTheorem}\end{equation}
 Clearly, $h'_1(E)=\Boltz\beta(E)$ and indeed $h_1(E)$ is invertible under our Postulate \ref{Postulate3} because for normal systems operator $H$ is unbounded from above and, therefore, $\beta$ is positive and also $S_{\rm vN,se}(E)$ is invertible.
\\
Now, select an initial stable equilibrium state for a system $B$ and one for a system $C$, and consider the pairs of processes adopted to define the function $f_{11}^{B\rightarrow C}$  in Lemma  \ref{Lemma3}.  In each such pair of processes, since the von Neumann entropies of $XB$ and  $XC$ do not change and the von Neumann entropy change of $X$ is fixed, it follows that the von Neumann entropy change of $C$ equals that of $B$, i.e., $S_{\rm vN,se}^C(E^C)-S_{\rm vN,se}^C(E^C_1)=S_{\rm vN,se}^B(E^B)-S_{\rm vN,se}^B(E^B_1)$ and so  the relation between the energies of the end (stable equilibrium) states of $B$ and $C$ is such that \begin{equation}h_1^C(E^C)=h_1^B(E^B)\ .\label{h1rel}\end{equation} Using the invertibility of  function $h_1(E)$, we may rewrite this relation as \begin{equation}E^C=  \ih_1^C(h_1^B(E^B))\ \label{QSMfunction1}.\end{equation} Equation (\ref{QSMfunction1}) shows that the single valued function defined by the pairs of processes defined in Lemma \ref{Lemma3} exists and is given by \begin{equation}E^C=  f_{11}^{B\rightarrow C}(E^B)=\ih_1^C(h_1^B(E^B))\ .\label{QSMfunction}\end{equation}
It is easily proved that $f_{11}^{B\rightarrow C}(E^B)$  is differentiable. In fact, by applying the chain rule and employing the identity (\ref{inverseTheorem}) and again Eq.\ (\ref{QSMfunction}), one obtains
\begin{equation}\frac{d \, f_{11}^{B\rightarrow C}(E^B)}{dE^B}= {\ih'}{}_1^C(h_1^B(E^B))\,{h'}{}_1^B(E^B) =\frac{{h'}{}_1^B(E^B)}{{h'}{}_1^C( \ih_1^C(h_1^B(E^B)))}=\frac{{h'}{}_1^B(E^B)}{{h'}{}_1^C( f_{11}^{B\rightarrow C}(E^B))}=\frac{{h'}{}_1^B(E^B)}{{h'}{}_1^C( E^C)}\ .\label{QSMderivative}\end{equation}
\\
Moreover, if instead of $B$ and $C$ we consider  systems $R$ and $B$ and we write the definition of temperature according to Eq.\ (\ref{temperature}) and use Eq.\ (\ref{QSMderivative}), we obtain
\begin{equation}\label{aaaa}\frac{T_1^B}{T_1^R}=\left.\frac{d\,  f_{11}^{R\rightarrow B}(E^R)}{dE^R}\right|_{E_1^R} =\frac{{h'}{}_1^R(E_1^R)}{{h'}{}_1^B( f_{11}^{R\rightarrow B}(E_1^R))}=\frac{{h'}{}_1^R(E_1^R)}{{h'}{}_1^B( E_1^B)}=\frac{\Boltz\beta^R(E_1^R)}{\Boltz\beta^B(E_1^B)}
 \end{equation}
 in agreement with the usual QSM identification $\beta(E_1)=1/\Boltz T_1$ for the temperature of the stable equilibrium state with energy $E_1$. From the continuity of $\beta(E)$, it follows that also our Assumption \ref{Assumption5} below is in agreement within QSM. \\
 Inserting the relation $dS_{\rm vN,se}(E)/dE=\Boltz\beta(E)$ and  the identification $\beta(E)=1/\Boltz T(E)$ in the integral calculus identity $\int_{E_1}^{E_2} \frac{dS_{\rm vN,se}(E)}{dE}\: dE= S_{\rm vN,se}(E_2)-S_{\rm vN,se}(E_1)$,  one obtains the relation $\int_{E_1}^{E_2} \frac{1}{T(E)}\:dE = S_{\rm vN,se}(E_2)-S_{\rm vN,se}(E_1)$ which substituted into  our new definition of entropy below, Eq.\ (\ref{entropyabs}), yields
 \begin{equation}
 S^A_1 = S^A_0   - \left[S^B_{\rm vN,se}\left(E^B_{\rm se2rev}\big|_{A_0 A_1}^{{\rm sw,}B_{\rm se1}}\right)-S^B_{\rm vN,se}\left(E^B_{\textrm{se}1}\right)\right] \;\; .
  \end{equation}
This relation, together with the additivity property of the von Neumann entropy functional and its invariance in unitary processes, essentially implies that our operational definition of entropy is fully compatible with the von Neumann entropy.
\end{comment}

\begin{coroll}\label{CorollaryTfunction}
The temperature of the stable equilibrium states of any system $B$ is a function of its energy $E^B$ and the region of space $\RB{}$  it occupies, i.e.,
\begin{equation}\label{TfunctionER}
T^B=T^B(E^B;\RB{}) \ ,
\end{equation}
provided the reference state $R_{\rm se1}$ of the reference system $R$ and the arbitrary multiplicative constant $T^R_{\rm se1}$ that are necessary for the definition of $T^B$ according to Eq.\ (\ref{temperature}) have been chosen once and for all. \dimostrazione{Corollary}{CorollaryTfunction}{The conclusion is a direct consequence of Postulate \ref{Postulate2}, Lemma \ref{Lemma1} and definition (\ref{temperature}).}
\end{coroll}

\begin{comment}
\textbf{Choice of  reference system and
reference stable equilibrium state.} In the macroscopic domain, the following choice of $R$
and of $R_{\rm se1}$ is currently employed, because it can be
easily reproduced in any laboratory. The reference system $R$ is
composed of a sufficient number of moles of pure water and its
reference stable equilibrium state $R_{\rm se1}$ is any of the
stable equilibrium states of $R$ in which ice, liquid water, and
water vapor coexist. This choice is convenient because, up to the
measurement accuracy available today, the value of the limit in
Eq.\ (\ref{temperature}) is practically independent of both the
number of moles in system $R$ and the particular choice of the
reference state $R_{\textrm{se}1}$, as long as it belongs to the
set of triple-point states.  With this selection for the reference stable equilibrium state, we obtain the S.I. thermodynamic
temperature, with unit called \textit{kelvin}, by setting
$T^R_{\textrm{se}1}=273.16$ K.\\
In the microscopic field, it could be convenient to choose as a reference system $R^\mu$ a few-particle monoatomic gas and as a reference state of $R^\mu$ the stable equilibrium state $R^\mu_{\textrm{se}1}$ which according to the standard model in quantum statistical mechanics has a temperature
$T^{R^\mu}_{\textrm{se}1}=273.16$ K. Note that, by the next theorem (Theorem \ref{Theorem4}), we  prove that the ratio of two temperatures can be measured directly and is independent of the choice of the reference system and of the reference stable equilibrium state. Hence, any system in any stable equilibrium state such that the temperature of the system is known can be used as a new reference system in a reference stable equilibrium state, without inconsistencies.
\end{comment}

\begin{thm}\label{Theorem4}
Let $B_{\rm se1}$ be any stable equilibrium state of a system $B$ and let $C_{\rm se1}$ be any stable equilibrium state of a system $C$, both with a non vanishing temperature. Then, the ratio of the temperatures of $B_{\rm se1}$ and $C_{\rm se1}$, as defined via Eq.\ (\ref{temperature}),
is independent of the choice of the reference system $R$ and of the reference stable equilibrium state $R_{\rm se1}$, and can be measured directly by the following procedure.\\
Consider the set of pairs of processes $\{(\Pi^{B_{\textrm{se}1}}_{XB\textrm{rev}}; \Pi^{C_{\textrm{se}1}}_{XC\textrm{rev}})\}$, where $\Pi^{B_{\textrm{se}1}}_{XB\textrm{rev}}$ is any reversible weight processes for $XB$ standard with respect $B$ and with initial state $B_{\textrm{se}1}$, $\Pi^{C_{\textrm{se}1}}_{XC\textrm{rev}}$ is any reversible weight processes for $XC$ standard with respect $C$, with initial state $C_{\textrm{se}1}$ and with the same initial and final state of $X$ as $\Pi^{B_{\textrm{se}1}}_{XB\textrm{rev}}$, and $X$ is a system which can be chosen and changed arbitrarily, as well as the initial state of $X$. On account of Lemma \ref{Lemma3} the set of pairs of processes $\{(\Pi^{B_{\textrm{se}1}}_{XB\textrm{rev}}; \Pi^{C_{\textrm{se}1}}_{XC\textrm{rev}})\}$ defines a single valued and invertible function $f^{B\rightarrow C}_{11}( E^B)$, which is differentiable in $E^B_{\textrm{se}1}$. The ratio of the temperatures $T^C_{\textrm{se}1}$ and $T^B_{\textrm{se}1}$ is given by
\begin{equation}\label{temperature-direct}
\!\frac{T^C_{\rm se1}}{T^B_{\rm se1}} \!=\!\!\! \lim_{E^B \rightarrow E^B_{\textrm{se}1}} \!\!\!\! \frac{f^{B\rightarrow C}_{11}\!( E^B) - f^{B\rightarrow C}_{11}\!( E^B_{\textrm{se}1})}{E^B - E^B_{\textrm{se}1}} \!=\!
\frac{\textrm{d}f^{B\rightarrow C}_{11}}{\textrm{d}E^B}\bigg|_{E^B_{\textrm{se}1}} \! .\!\!
\end{equation}
 \dimostrazione{Theorem}{Theorem4}{By applying to Eq.\ (\ref{function-function}) the theorem on the derivative of a composite function, one obtains
\begin{equation}\label{derivative-composite}
\frac{\textrm{d} f^{B\rightarrow C}_{11}}{\textrm{d} E^B}\bigg|_{E^B_{\textrm{se}1}} \!\!= \frac{\textrm{d} f^{R\rightarrow C}_{11}}{\textrm{d}  E^R}\bigg|_{E^R=f^{B\rightarrow R}_{11}( E^B_{\textrm{se}1})} \; \frac{\textrm{d} f^{B\rightarrow R}_{11}}{\textrm{d} E^B}\bigg|_{E^B_{\textrm{se}1}} .
\end{equation}
On account of Eq.\ (\ref{temperature}), the first derivative at the right hand side of Eq.\ (\ref{derivative-composite}) can be rewritten as
\begin{equation}\label{derivative-composite-first}
 \frac{\textrm{d} f^{R\rightarrow C}_{11}}{\textrm{d} E^R}\bigg|_{E^R_{\textrm{se}1}} = \frac{T^C_{\rm se1}}{T^R_{\rm se1}} \;\; .
\end{equation}
By applying Eqs.\ (\ref{derivative-inverse}) and (\ref{temperature}), the second derivative at the right hand side of Eq.\ (\ref{derivative-composite}) can be rewritten as
\begin{equation}\label{derivative-composite-second}
\frac{\textrm{d} f^{B\rightarrow R}_{11}}{\textrm{d} E^B}\bigg|_{E^B_{\textrm{se}1}} = \frac{1}{\displaystyle \frac{\textrm{d} f^{R\rightarrow B}_{11}}{\textrm{d} E^R}\bigg|_{E^R_{\textrm{se}1}}} = \frac{1}{\displaystyle \frac{T^B_{\rm se1}}{T^R_{\rm se1}}} = \frac{T^R_{\rm se1}}{T^B_{\rm se1}} \;\; .
\end{equation}
By combining Eqs.\ (\ref{derivative-composite}), (\ref{derivative-composite-first}) and (\ref{derivative-composite-second}) we obtain Eq.\ (\ref{temperature-direct}).} Theorem \ref{Theorem4} completes the definition of  temperature of a stable equilibrium state.
\end{thm}

\section{\label{EntropyDef}Definition of thermodynamic entropy for any state}

\begin{coroll}\label{Corollarydf22}
Consider a pair of stable equilibrium states $(B_{\textrm{se}1}, C_{\textrm{se}1})$ and the set of processes which defines the function  $E^C=f^{B\rightarrow C}_{11}(E^B)$ according to Lemma \ref{Lemma3}. Then, for every pair of stable equilibrium states of $B$ and $C$ determined by the same regions of space $\RB{}$ and $\RC{}$ as $B_{\textrm{se}1}$ and $C_{\textrm{se}1}$, respectively, and by the energy values $E^B$ and $E^C=f^{B\rightarrow C}_{11}(E^B)$,
\begin{equation}\label{new-equality22}
\frac{T^C\!(E^C\!\!=\! f^{B\rightarrow C}_{11}\!(E^B);\RC{})}{T^B\!(E^B;\RB{})} = \left.\frac{\textrm{d}f^{B\rightarrow C}_{11}\!(E^B)}{\textrm{d}E^B}\right|_{E^B}\ .
\end{equation}
 \dimostrazione{Corollary}{Corollarydf22}{For the fixed regions of space $\RB{}$, consider the set of stable equilibrium states  of system $B$ defined by varying the energy $E^B$. Select a value of energy $E^B$ and denote the corresponding  state in this set by $B_{\textrm{se}2}$, i.e., $E^B_{\textrm{se}2}=E^B$. Consider the pair of stable equilibrium states $(B_{\textrm{se}2}, C_{\textrm{se}2})$, where $C_{\textrm{se}2}$ is such that $E^C_{\textrm{se}2} = f^{B\rightarrow C}_{11}(E^B_{\textrm{se}2})$ and let $E^C=f^{B\rightarrow C}_{22}(E^B)$ be the function defined according to Lemma \ref{Lemma3}. Then, we have
\begin{equation}\label{new-equality33}
\frac{T^C\!(E^C_{\textrm{se}2};\RC{})}{T^B\!(E^B_{\textrm{se}2};\RB{})} \!=\!\frac{T^C_{\textrm{se}2}}{T^B_{\textrm{se}2}} \!=\! \left.\frac{\textrm{d}f^{B\rightarrow C}_{22}\!(E^B)}{\textrm{d}E^B}\right|_{E^B_{\textrm{se}2}} \!\!\!=\! \left.\frac{\textrm{d}f^{B\rightarrow C}_{11}\!(E^B)}{\textrm{d}E^B}\right|_{E^B_{\textrm{se}2}} \!,
\end{equation}
where the first equality obtains from Eq.\ (\ref{TfunctionER}), the second   from Eq.\ (\ref{temperature-direct}) applied to $f^{B\rightarrow C}_{22}(E^B)$, and the third  from  Eq.\ (\ref{new-equality44}).
Recalling that $E^B_{\textrm{se}2}=E^B$, that $E^B$ can be varied arbitrarily, and that $E^C_{\textrm{se}2}=f^{B\rightarrow C}_{11}(E^B)$,  Eq.\ (\ref{new-equality33}) yields Eq.\ (\ref{new-equality22}).}
\end{coroll}

\begin{assumption}\label{Assumption5} For every system $B$ and every choice of the regions of space $\RB{}$ occupied by the constituents of $B$, the temperature of the stable equilibrium states of $B$ is a continuous function of the energy of $B$ and is vanishing only in the stable equilibrium state with the lowest energy for the given regions of space $\RB{}$.
\end{assumption}

\begin{lmm}\label{Lemma5}
For every pair of stable equilibrium states $B_{\textrm{se}1}$ and $B_{\textrm{se}2}$ of a system $B$, with a non vanishing temperature and with the same regions of space $\RB{}$ occupied by the constituents of $B$, the integral
\begin{equation}\label{integral-A}
\int_{E^B_{\textrm{se}1}}^{E^B_{\textrm{se}2}}
\frac{1}{T^B(E^B;\RB{})} \; \textrm{d}E^B \;\; ,
\end{equation}
 has a finite value and the same sign as $E^B_{\textrm{se}2} - E^B_{\textrm{se}1}$. \dimostrazione{Lemma}{Lemma5}{Since both $E^B_{\textrm{se}1}$ and $E^B_{\textrm{se}2}$ are greater than the lowest energy value for the given regions of space $\RB{}$, on account of Assumption \ref{Assumption5} the function $1 /\,T^B(E^B;\RB{})$ is defined and continuous in the whole interval. Therefore the integral in Eq.\ (\ref{integral-A}) exists. Moreover, on account of Corollary \ref{Corollaryb}, the function $1 /\,T^A(E;\RA{})$ has positive values. Therefore, if $E^A_{\textrm{se}2} > E^A_{\textrm{se}1}$ the integral in Eq.\ (\ref{integral-A}) has a positive value; if $E^A_{\textrm{se}2} < E^A_{\textrm{se}1}$ the integral in Eq.\ (\ref{integral-A}) has a negative value.}
 \end{lmm}

\begin{thm}\label{Theorem5}
Consider an arbitrarily chosen pair of states  $(A_1, A_2)$ of a system $A$, such that $A$ is separable and uncorrelated from its environment, another system $B$ in the environment of $A$ and a
reversible weight process $\Pi^{B_{\textrm{se}1}}_{AB\rm rev}$ for $AB$ in which $A$ goes from $A_1$ to $A_2$, standard with respect to $B$ and with initial state $B_{\textrm{se}1}$, chosen so that the temperature of $B$ is non vanishing both for $B_{\textrm{se}1}$ and for the final state $B_{\textrm{se}2}$. Denote by $\RB{}$ the regions of space occupied by the constituents of $B$ in its end states $B_{\textrm{se}1}$ and
$B_{\textrm{se}2}$. Then the value of the integral
\begin{equation}\label{integral}
\int_{E^B_{\textrm{se}1}}^{E^B_{\rm se2rev}\big|_{A_1 A_2}^{{\rm sw,}B_{\rm se1}}}
\frac{1}{T^B(E^B;\RB{})} \; \textrm{d}E^B \;\; ,
\end{equation}
depends only on the
pair of states $(A_1, A_2)$ of system $A$ and is independent of the choice of
system $B$, of the initial stable equilibrium state
$B_{\textrm{se}1}$, and of the details of the reversible
weight process for $AB$, standard with respect to $B$.  \dimostrazione{Theorem}{Theorem5}{On account of Theorem \ref{Theorem2}, once the initial state $B_{\textrm{se}1}$ has been chosen, the final state $B_{\textrm{se}2}$ is determined uniquely. Therefore, the value of the integral in Eq.\ (\ref{integral}) can depend, at most, on the pair of states $(A_1, A_2)$ and on the choice of system $B$ and of its initial state $B_{\textrm{se}1}$.
Consider another system $C$ and a reversible weight process $\Pi^{C_{\textrm{se}1}}_{AC\rm rev}$ for $AC$ in which $A$  goes again from $A_1$ to $A_2$, standard with respect to $C$ and with an initial state $C_{\textrm{se}1}$ chosen arbitrarily, provided that the temperature of $C$ is non vanishing both for $C_{\textrm{se}1}$ and for the final state $C_{\textrm{se}2}$. We will prove that the integral
\begin{equation}\label{integral-C}
\int_{E^C_{\textrm{se}1}}^{E^C_{\rm se2rev}\big|_{A_1 A_2}^{{\rm sw,}C_{\rm se1}}}
\frac{1}{T^C(E^C;\RC{})} \; \textrm{d}E^C \;\;
\end{equation}
has the same value as the integral in Eq.\ (\ref{integral}), implying that such value is independent of the choice of system $B$ and of the initial state $B_{\textrm{se}1}$, and, therefore, it depends only on the pair of states $(A_1, A_2)$.\\
The set of pairs of processes $\{(\Pi^{B_{\textrm{se}1}}_{AB\rm rev},\Pi^{C_{\textrm{se}1}}_{AC\rm rev})\}$ such that the energy of the final state of $B$ is in the range $E^B_{\textrm{se}1}\le E^B \le E^B_{\textrm{se}2} $  belongs to the set defined in Lemma \ref{Lemma3}, so that $E^C=f_{11}^{B\rightarrow C} (E^B)$ and, since this function is invertible (Lemma \ref{Lemma3}), $E^B=f_{11}^{C\rightarrow B} (E^C)$ so that, in particular, $E^B_{\textrm{se}1}=f_{11}^{C\rightarrow B} (E^C_{\textrm{se}1})$ and  $E^B_{\rm se2rev}\big|_{A_1 A_2}^{{\rm sw,}B_{\rm se1}}=f_{11}^{C\rightarrow B} \big(E^C_{\rm se2rev}\big|_{A_1 A_2}^{{\rm sw,}C_{\rm se1}}\big)$. Now, consider the change of integration variable in the definite integral (\ref{integral-C}) from $E^C=f_{11}^{B\rightarrow C} (E^B)$ to $E^B$. By virtue of Eq.\ (\ref{new-equality22}) (Corollary \ref{Corollarydf22}) we have
\begin{equation}\label{new-equality1}
    \textrm{d}E^C\!=\!\frac{\textrm{d}f^{B\rightarrow C}_{11}(E^B)}{\textrm{d}E^B}\;\textrm{d}E^B \!=\! \frac{T^C(f_{11}^{B\rightarrow C} (E^B);\RC{})}{T^B(E^B;\RB{})}\;\textrm{d}E^B  .
\end{equation}
Thus, the integral in Eq.\ (\ref{integral-C}) can be rewritten as follows
\begin{eqnarray}\label{integral-equality}
&&\!\!\!\!\!\!\!\!\!\!\!\int_{f_{11}^{C\rightarrow B} (E^C_{\textrm{se}1})}^{f_{11}^{C\rightarrow B} \big(E^C_{\rm se2rev}\big|_{A_1 A_2}^{{\rm sw,}C_{\rm se1}}\big)}
\frac{1}{T^C(f_{11}^{B\rightarrow C} (E^B);\RC{})} \frac{T^C(f_{11}^{B\rightarrow C} (E^B);\RC{})}{T^B(E^B;\RB{})}\;\textrm{d}E^B \nonumber\\ &&\!\!\!\!\!\!\!\!\!\!\!= \int_{E^B_{\textrm{se}1}}^{E^B_{\rm se2rev}\big|_{A_1 A_2}^{{\rm sw,}B_{\rm se1}}}
\frac{1}{T^B(E;\RB{})} \; \textrm{d}E^B  \end{eqnarray}}
\end{thm}

\begin{definition}
\textbf{Definition of thermodynamic entropy.} Let ($A_1$,$A_2$) be any pair of states of a
system $A$, such that $A$ is separable and uncorrelated from its environment,
and let $B$ be any other system placed in the
environment of $A$. We call \textit{entropy difference} between
$A_2$ and $A_1$ the quantity
\begin{equation}\label{entropy}
S^A_2 - S^A_1 = - \int_{E^B_{\textrm{se}1}}^{E^B_{\rm se2rev}\big|_{A_1 A_2}^{{\rm sw,}B_{\rm se1}}}
\frac{1}{T^B(E^B;\RB{})} \; \textrm{d}E^B \;\; ,
\end{equation}
where $B_{\textrm{se}1}$ and $B_{\textrm{se}2}$ are the initial
and the final state of $B$ in any reversible weight process for
$AB$ from $A_1$ to $A_2$, standard with respect to $B$, $\RB{}$ is
the set of regions of space occupied by the constituents of $B$ in
the states $B_{\textrm{se}1}$ and $B_{\textrm{se}2}$, and $T^B$ is the
temperature of $B$. The initial state $B_{\textrm{se}1}$ is chosen so that both $T^B_{\textrm{se}1}$ and $T^B_{\textrm{se}2}$ are non vanishing.
On account of Theorem \ref{Theorem5}, the right hand side
of Eq.\ (\ref{entropy}) is determined uniquely by states $A_1$ and
$A_2$.\\ Let $A_0$ be a reference state of $A$, to which we assign
an arbitrarily chosen value of entropy $S^A_0$. Then, the value of
the entropy of $A$ in any other state $A_1$ of $A$ such that $A$ is separable and uncorrelated from its environment is determined uniquely by the equation
\begin{equation}\label{entropyabs}
S^A_1 = S^A_0  - \int_{E^B_{\textrm{se}1}}^{E^B_{\rm se2rev}\big|_{A_0 A_1}^{{\rm sw,}B_{\rm se1}}}
\frac{1}{T^B(E;\RB{})} \; \textrm{d}E^B \;\; ,
\end{equation}
where $B_{\textrm{se}1}$ and $B_{\textrm{se}2}$ are the initial
and the final state of $B$ in any reversible weight process for
$AB$ from $A_0$ to $A_1$, standard with respect to $B$, $T^B_{\textrm{se}1}$ and $T^B_{\textrm{se}2}$ are non vanishing, and the
other symbols have the same meaning as in Eq.\ (\ref{entropy}).
Such a process exists for every state $A_1$ such that $A$ is separable and uncorrelated from its environment, in a set of states where Assumption \ref{Assumption3} holds.
\end{definition}

\begin{lmm}\label{Lemma7}  Let ($A_1$, $A_2$) be any pair of
states of a system $A$ such that $A$ is separable and uncorrelated from its environment, and let $B$ be any other system placed in the environment of $A$. Let $\Pi_{AB\rm irr}$ be any irreversible weight process for $AB$, standard with respect to $B$, from $A_1$ to $A_2$, and let $B_{\textrm{se}1}$ and $B_{\textrm{se}2\textrm{irr}}$ be the end states of $B$ in the process. Then
\begin{equation}\label{inequality}
- \int_{E^B_{\textrm{se}1}}^{E^B_{\rm se2irr}\big|_{A_1 A_2}^{{\rm sw,}B_{\rm se1}}}
\frac{1}{T^B(E^B;\RB{})} \; \textrm{d}E^B < S^A_2 - S^A_1 \;\; .
\end{equation}
 \dimostrazione{Lemma}{Lemma7}{Let $\Pi_{AB\rm rev}$ be any reversible weight process
for $AB$, standard with respect to $B$, from $A_1$ to $A_2$, with the same initial state $B_{\textrm{se}1}$ of $B$, and let $B_{\textrm{se}2\textrm{rev}}$ be the final state of $B$ in this process. On
account of Theorem \ref{Theorem2},
\begin{equation}\label{minimum-energychange}
E^B_{\rm se2rev}\big|_{A_1 A_2}^{{\rm sw,}B_{\rm se1}} < E^B_{\rm se2irr}\big|_{A_1 A_2}^{{\rm sw,}B_{\rm se1}} \;\; .
\end{equation}
Since $T^B$ is a positive function, from Eqs.\ (\ref{minimum-energychange})
and (\ref{entropy}) one obtains
\begin{equation}\label{inequality-two}
\!- \!\!\int_{E^B_{\textrm{se}1}}^{E^B_{\rm se2irr}\big|_{A_1 A_2}^{{\rm sw,}B_{\rm se1}}} \!
\frac{1}{T^B(E^B;\RB{})} \; \textrm{d}E^B \! <\!  - \!\! \int_{E^B_{\textrm{se}1}}^{E^B_{\rm se2rev}\big|_{A_1 A_2}^{{\rm sw,}B_{\rm se1}}} \!\!
\frac{1}{T^B(E^B;\RB{})} \; \textrm{d}E^B \!= \! S^A_2\! -\! S^A_1.
\end{equation}}
\end{lmm}

\begin{comment}
\textbf{Approximate measurement of the entropy difference.}
A conceptually different but practically important issue is that, even within the domain of validity of Assumption \ref{Assumption3} in a given nonequilibrium framework, the reversible weight processes for  $AB$, standard with respect to $B$, that are assumed to be conceivable,  may nevertheless be difficult or practically impossible to implement, whereas it may be easier to approximate them irreversibly.
\\
Consider  a closed system $A$ and a system $B$ in its environment that fulfill Assumption \ref{Assumption3} for the pair of states ($A_1$, $A_2$) of $A$ and state $B_{\textrm{se}1}$ of $B$, as well as for the pair of states ($A_2$, $A_1$) of $A$ and state $B_{\textrm{se}1'}$ of $B$. Assume that for both pairs we can perform an irreversible process under the conditions of Lemma \ref{Lemma7}.
In such cases, of course, system $B$ ends in different states than if the process were reversible: we denote the energies of the end states of $B$ by $E^B_{\rm se2irr}\big|_{A_1 A_2}^{{\rm sw,}B_{\rm se1}}$ and $E^B_{\rm se2'irr}\big|_{A_2 A_1}^{{\rm sw,}B_{\rm se1'}}$, respectively.
Then, applying  Lemma \ref{Lemma7} to the two processes we obtain
%$- \int_{E^B_{\textrm{se}1}}^{E^B_{\rm se2irr}\big|_{A_1 A_2}^{{\rm sw,}B_{\rm se1}}} \frac{1}{T^B(E^B;\RB{})} \; \textrm{d}E^B < S^A_2 - S^A_1 $ and $- \int_{E^B_{\textrm{se}1'}}^{E^B_{\rm se2'irr}\big|_{A_2 A_1}^{{\rm sw,}B_{\rm se1'}}} \frac{1}{T^B(E^B;\RB{})} \; \textrm{d}E^B < S^A_1 - S^A_2$. Therefore,
\begin{equation}\label{approximate}
- \int_{E^B_{\textrm{se}1}}^{E^B_{\rm se2irr}\big|_{A_1 A_2}^{{\rm sw,}B_{\rm se1}}}
\frac{1}{T^B(E^B;\RB{})} \; \textrm{d}E^B < S^A_2 - S^A_1 < \int_{E^B_{\textrm{se}1'}}^{E^B_{\rm se2'irr}\big|_{A_2 A_1}^{{\rm sw,}B_{\rm se1'}}} \frac{1}{T^B(E^B;\RB{})} \; \textrm{d}E^B  \;\; .
\end{equation}
Equation \ref{approximate} allows one to establish experimentally, by performing two irreversible processes (protocols), an upper and a lower bound to the value of the entropy difference $S^A_2 - S^A_1$. By repeated experiments it is thus possible to restrict the range between lower and upper bound so as to converge towards an acceptable degree of approximation.
\end{comment}

\section{Principle of entropy non-decrease, additivity of entropy, maximum entropy principle}

Based on the above construction, in this section we obtain some of the main standard theorems about entropy and entropy change.

\begin{thm}\label{Theorem6} \textbf{Principle of entropy non-decrease in weight processes}.
Let $(A_1, A_2)$ be a pair of  states of a system $A$ such that $A$ is separable and uncorrelated from its environment and let $A_1\xrightarrow{\rm w}A_2$ be any weight process for $A$ from $A_1$
to $A_2$. Then, the entropy difference $S^A_2 - S^A_1$ is equal to
zero if and only if the weight process is reversible; it is
strictly positive if and only if the weight process is
irreversible. \dimostrazione{Theorem}{Theorem6}{If  $A_1\xrightarrow{\rm w}A_2$ is
reversible, then it is a special case of a reversible
weight process for $AB$, standard with respect to $B$, in which the initial stable equilibrium
state of $B$ does not change. Therefore, $E^B_{\rm se2rev}\big|_{A_1 A_2}^{{\rm sw,}B_{\rm se1}} = E^B_{\textrm{se}1}$ and  Eq.\ (\ref{entropy}) yields
\begin{equation}\label{nondecrease-rev}
S^A_2 - S^A_1 = - \int_{E^B_{\textrm{se}1}}^{E^B_{\rm se2rev}\big|_{A_1 A_2}^{{\rm sw,}B_{\rm se1}}= E^B_{\textrm{se}1}}
\frac{1}{T^B(E^B;\RB{})} \; \textrm{d}E^B = 0 \;\; .
\end{equation}
If  $A_1\xrightarrow{\rm w}A_2$ is irreversible, then it is a
special case of an irreversible weight process for $AB$, standard with respect to $B$,
in which the initial stable equilibrium state of $B$ does not
change. Therefore, $E^B_{\rm se2irr}\big|_{A_1 A_2}^{{\rm sw,}B_{\rm se1}} = E^B_{\textrm{se}1}$ and Eq.
(\ref{inequality})  yields
\begin{equation}\label{nondecrease-irr}
S^A_2 - S^A_1 > - \int_{E^B_{\textrm{se}1}}^{E^B_{\rm se2irr}\big|_{A_1 A_2}^{{\rm sw,}B_{\rm se1}}=E^B_{\textrm{se}1}}
\frac{1}{T^B(E^B;\RB{})} \; \textrm{d}E^B = 0 \;\; .
\end{equation}
Moreover, if a weight process $A_1\xrightarrow{\rm w}A_2$ for
$A$ is such that $S^A_2 - S^A_1 = 0$, then the process must be
reversible, because we just proved that for any irreversible
weight process  $S^A_2 - S^A_1 > 0$; if a weight process $A_1\xrightarrow{\rm w}A_2$ for $A$ is such that $S^A_2 - S^A_1 > 0$, then
the process must be irreversible, because we just proved that for
any reversible weight process $S^A_2 - S^A_1 = 0$.}
\end{thm}

\begin{thm}\label{Theorem7} \textbf{ Additivity of entropy differences}.
Consider the pair of states $(C_1 = A_1B_1, C_2 = A_2 B_2)$ of the
composite system $C =A B$, such that $A$, $B$ and $C$ are separable and uncorrelated from their environment. Then,
\begin{equation}\label{entropyadditivity}
S^{AB}_{ 2,2 } - S^{AB}_{ 1,1 } = S^A_2 - S^A_1 + S^B_2 -
S^B_1 \;\; .
\end{equation}
 \dimostrazione{Theorem}{Theorem7}{Let us choose a system $D$ (with fixed
regions of space $\RD{}$) in the environment of $C$, and consider the processes  $\Pi_{AD\rm
rev}=A_1D_{\rm se1}\xrightarrow{\rm wrev}A_2D_{\rm se3rev}$ and
$\Pi_{BD\rm rev}=B_1D_{\rm se3rev}\xrightarrow{\rm wrev}B_2D_{\rm
se2rev}$. For process $\Pi_{AD\rm rev}$ Eq.\ (\ref{entropy}) implies that
\begin{equation}\label{Proof7-A}
S^A_2 - S^A_1 =- \int_{E^D_{\textrm{se}1}}^{E^D_{\rm se3rev}\big|_{A_1 A_2}^{{\rm sw,}D_{\rm se1}}}
\frac{1}{T^D(E^D;\RD{})} \; \textrm{d}E^D\ .\end{equation} For process
$\Pi_{BD\rm rev}$ Eq.\ (\ref{entropy}) implies that
\begin{equation}\label{Proof7-B}S^B_2 - S^B_1 =- \int_{E^D_{\rm se3rev}\big|_{A_1 A_2}^{{\rm sw,}D_{\rm se1}}}^{E^D_{\rm se2rev}\big|_{B_1 B_2}^{{\rm sw,}D_{\rm se3rev}}}
\frac{1}{T^D(E^D;\RD{})} \; \textrm{d}E^D\ .\end{equation} The composite process
$(\Pi_{AD\rm rev},\Pi_{BD\rm rev}) = A_1B_1D_{\rm
se1} \xrightarrow{\rm wrev} A_2B_1D_{\rm se3rev} \xrightarrow{\rm
wrev} A_2B_2D_{\rm se2rev}$ is a reversible weight process from
$C_1=A_1B_1$ to $C_2=A_2 B_2$ for $CD$, standard with respect to
$D$, in which the energy change of $D$ is the sum of its energy
changes in the constituent processes $\Pi_{AD\rm rev}$ and
$\Pi_{BD\rm rev}$. Therefore,  Eq.\ (\ref{entropy}) implies that
\begin{equation}\label{Proof7-C}S^C_2 - S^C_1 =- \int_{E^D_{\textrm{se}1}}^{E^D_{\rm se2rev}\big|_{C_1 C_2}^{{\rm sw,}D_{\rm se1}}=E^D_{\rm se3rev}\big|_{A_1 A_2}^{{\rm sw,}D_{\rm se1}}+E^D_{\rm se2rev}\big|_{B_1 B_2}^{{\rm sw,}D_{\rm se3rev}}}
\frac{1}{T^D(E^D;\RD{})} \; \textrm{d}E^D\ .\end{equation} Subtracting Eqs.\
(\ref{Proof7-A}) and  (\ref{Proof7-B}) from Eq.\ (\ref{Proof7-C})
%and using $E^D_{\rm se2rev}\big|_{C_1 C_2}^{{\rm sw,}D_{\rm se1}}=E^D_{\rm se2rev}\big|_{A_1 A_2}^{{\rm sw,}D_{\rm se1}}+E^D_{\rm se2rev}\big|_{B_1 B_2}^{{\rm sw,}D_{\rm se3rev}}$
yields Eq.\ (\ref{entropyadditivity}).}
\end{thm}

\begin{comment} As a consequence of Theorem \ref{Theorem7}, if the
values of entropy are chosen so that they are additive in the
reference states, entropy results as an additive property.\end{comment}

\spazio \begin{thm}\label{Theorem8}\textbf{Maximum entropy principle}. Consider a closed system $A$, and the set of all the states of $A$ with a given value $E^A_1$ of the energy, given regions of space $\RA{}$, and
such that $A$ is separable and uncorrelated from its environment.
Then, the entropy of $A$ has the highest value in this set of states only in the unique
stable equilibrium state $A_{\rm se1}=A_{\rm se}(E^A_1;\RA{})$
determined by $\RA{}$ and the value $E^A_1$. \dimostrazione{Theorem}{Theorem8}{Let $A_1$ be any state different from $A_{\rm se1}$  in the set of states of $A$ considered here. On account of Postulate \ref{Assumption2}  a zero work weight process
$\W{A}{1}{{\rm se1}}$ exists and is irreversible because a zero work weight process $\W{A}{{\rm se1}}{1}$
would violate the definition of stable equilibrium state.
Therefore,  Lemma \ref{Lemma7} implies $S^A_{\rm se1} >
S^A_1 $.}
\end{thm}

\section{ Extension of the definition of entropy to sets of states that do not fulfill Assumption 1 }

 In this section we extend our operational definition of thermodynamic entropy to a broader class of system models for which Assumption 1 is fulfilled only within one or more subsets of states. We begin by relaxing Assumption 1 as follows.

\begin{assumptionrelaxed}Any given state $A_1$ of any closed system $A$ such that $A$ is separable and uncorrelated from its environment, either belongs to a set $\Sigma_A$ where every pair of states fulfills Assumption 1 or it can be reached from at least one state of $\Sigma_A$ by means of a weight process for $A$ and it can be the initial state of at least one weight process for A having as final state a state of $\Sigma_A$.
\end{assumptionrelaxed}

\begin{definition} \textbf{Definition of thermodynamic entropy.}
For the states in $\Sigma_A$ we adopt the definition given in Section {\ref{EntropyDef}}. For every state $A_1$ that does not belong to $\Sigma_A$, we associate a range of entropy values, as follows.  Let $A_{\rm 1low}$ be the state with highest entropy, in $\Sigma_A$, such that a weight process for $A$ from $A_{\rm 1low}$ to $A_1$ is possible, and let $A_{\rm 1high}$ be the state with lowest entropy, in $\Sigma_A$, such that a weight process for $A$ from $A_{\rm 1high}$ to $A_1$ is possible. The existence of such states is granted by Relaxed Assumption 1. Then the range of entropy values associated with state $A_1$ is
\begin{equation}\label{entropyrange}
S^A_{\rm 1low}\le S^A_1\le S^A_{\rm 1high} \;\; .
\end{equation}
\end{definition}

\begin{thm}\label{Theorem9} \textbf{Principle of entropy non-decrease in weight processes}.
Let $A_1$ and $A_2$ be two states of system $A$, such that all the entropy values in the range associated with $A_2$ are higher than all the entropy values in the range associated with $A_1$, namely, $S^A_{\rm 2low}>S^A_{\rm 1high}$. Then a weight process for $A$ from  $A_2$ to  $A_1$ is impossible.
\dimostrazione{Theorem}{Theorem9}{ Weight processes $A_{\rm 2low}\xrightarrow{\rm w}A_2$ and $A_1\xrightarrow{\rm w}A_{\rm 1high}$  exist by definition of $A_{\rm 2low}$ and of $A_{\rm 1high}$. Suppose that, contrary to the conclusion, a weight process $A_2\xrightarrow{\rm w}A_1$  exists.
Then, a weight process  $A_{\rm 2low}\xrightarrow{\rm w}A_2\xrightarrow{\rm w}A_1\xrightarrow{\rm w}A_{\rm 1high}$ would exist and, since $S^A_{\rm 2low}>S^A_{\rm 1high}$, would violate the principle of entropy nondecrease in $\Sigma_A$ already proved in Theorem \ref{Theorem6}. }\end{thm}

%\begin{comment}
%Relaxed Assumption 1 can be further relaxed to contemplate: (i) systems with several (obviously disjoint) sets of states $\Sigma^n_A$ where every pair of states fulfills Assumption 1; and (ii) a state that does not belong to any one of these sets is an intermediate state of at least a composite weight process for $A$ that connects either two states of the same set $\Sigma^n_A$ or a state of $\Sigma^m_A$ with one of $\Sigma^n_A$. For the states in each set $\Sigma^n_A$ we adopt the definition given in Section \ref{EntropyDef}. Clearly, in order to achieve a principle of entropy non-decrease in all weight processes the reference values $S^A_{0n}$ in Eq.\  (\ref{entropyabs}) must be suitably chosen so that the highest value of the entropy in $\Sigma^m_A$ is smaller than the lowest in $\Sigma^n_A$ if there is at least a state $A_1$ that is an intermediate state of a composite weight process for $A$ that starts from a state in $\Sigma^m_A$ and ends in one in $\Sigma^n_A$.
%\end{comment}

\begin{thm}\label{Theorem10}\textbf{Additivity of entropy}. Consider a state $(AB)_1 =A_1B_1$ of a composite system $AB$ fulfilling Relaxed Assumption 1 and such that $A$ and $B$ are separable and uncorrelated. Denote by $ \left[ S^A_{\rm 1low},S^A_{\rm 1high}\right]$ the range of entropy values associated with state $A_1$, and with $ \left[ S^B_{\rm 1low},S^B_{\rm 1high}\right]$ the range for state $B_1$. If the entropy values in the reference states of $A$, $B$ and $AB$  have been chosen so that in $\Sigma_{AB}$ the entropy of any state of $AB$ in which $A$ and $B$ are separable and uncorrelated equals the sum of  the entropy values of A and B, then the range of entropy values associated with state $A_1B_1$ is contained in the interval $ \left[  S^A_{\rm 1low}+S^B_{\rm 1low},S^A_{\rm 1high}+S^B_{\rm 1high}\right]$.
\dimostrazione{Theorem}{Theorem10}{ Let $(AB)_{\rm 1low}$ be the highest entropy state in $\Sigma_{AB}$ such that a weight process for $AB$ from $(AB)_{\rm 1low}$ to  $A_1B_1$ is possible. A weight process for $AB$ from state $A_{\rm 1low}B_{\rm 1low}$ to $A_1B_1$ is possible, because it can be obtained by two separate weight processes for $A$ and $B$. Therefore,  $S^{AB}_{\rm 1low,1low}\le S^{AB}_{\rm 1low}$ or, using the additivity in $\Sigma_A\times\Sigma_B$,
\begin{equation}\label{proof91}
S^A_{\rm 1low}+S^B_{\rm 1low}=S^{AB}_{\rm 1low,1low} \le S^{AB}_{\rm 1low} \;\; .
\end{equation}
Similarly, let $(AB)_{\rm 1high}$ be the lowest entropy state in $\Sigma_{AB}$ such that a weight process for $AB$ from $A_1B_1$ to $(AB)_{\rm 1high}$  is possible. A weight process for $AB$ from state $A_1B_1$ to $A_{\rm 1high}B_{\rm 1high}$ is possible, because it can be obtained by two separate weight processes for $A$ and $B$. Thus, we have  $S^{AB}_{\rm 1high}\le S^{AB}_{\rm 1high,1high}$, i.e.,
 \begin{equation}\label{proof92}
S^{AB}_{\rm 1high}\le S^{AB}_{\rm 1high,1high}= S^A_{\rm 1high}+S^B_{\rm 1high}\;\; .
\end{equation}
By definition, the entropy range associated with state  $(AB)_1$ is
 \begin{equation}\label{proof93}
S^{AB}_{\rm 1low}\le S^{AB}_1 \le S^{AB}_{\rm 1high}\;\; .
\end{equation}
Therefore, combining Eqs.\ (\ref{proof91}), (\ref{proof92}), (\ref{proof93}) yields our conclusion
 \begin{equation}\label{proof93}
S^A_{\rm 1low}+S^B_{\rm 1low}=S^{AB}_{\rm 1low,1low} \le S^{AB}_{\rm 1low}\le S^{AB}_1 \le S^{AB}_{\rm 1high}\le S^{AB}_{\rm 1high,1high}= S^A_{\rm 1high}+S^B_{\rm 1high}\;\; .
\end{equation}
 }\end{thm}

\section{Conclusions}

We presented a rigorous and general logical construction of an operational definition of thermodynamic entropy which can be applied, \emph{in principle}, even to non-equilibrium states of few-particle systems, provided they are separable and uncorrelated from their environment. The new logical construction provides an operational definition of  entropy which requires neither the concept of \emph{heat} nor that of \emph{thermal reservoir}. Therefore, it removes: (1) the logical   limitations that restrict \emph{a priori} the traditional definitions of entropy to the equilibrium states of many-particle systems; and (2) the operational limitations that restrict \emph{in practice}  to many-particle systems our previous definitions of non-equilibrium entropy because they hinged on the notion of thermal reservoirs.

\bibliographystyle{srt}
%
%%\nocite{*}
%
%%\hfill \today
%
%\bibliography{BerettaZanchniPeloritanav2bib}
%
%\end{document}

\end{document}